\documentstyle[preprint,aps]{revtex}
\begin{document}
\draft
\renewcommand{\theequation}{\arabic{section}.\arabic{equation}}
\title{Spatial structure of anomalously localized states in disordered
conductors.}
\author{ Alexander D. Mirlin}
\address{Institut f\"{u}r Theorie der Kondensierten Materie,
  Universit\"{a}t Karlsruhe, 76128 Karlsruhe, Germany}
\address{
and  Petersburg Nuclear Physics Institute, 188350 Gatchina, St.Petersburg,
Russia.}
\date{\today}
\maketitle
%\narrowtext
\tighten
\begin{abstract}
The spatial structure of wave functions of anomalously localized
states (ALS) in disordered conductors is studied in the framework of the
$\sigma$--model approach. These states are responsible for slowly
decaying tails of various distribution functions. In the
quasi-one-dimensional case, properties of ALS governing the asymptotic
form of the distribution of eigenfunction amplitudes are investigated
with the use of the transfer matrix method, which yields an exact
solution to the problem. Comparison of the results with those obtained
in the saddle-point approximation to the problem shows that the
saddle-point configuration correctly describes the smoothed intensity
of an ALS. On this basis, the properties of ALS in higher spatial
dimensions are considered. We study also the ALS responsible for the
asymptotic behavior of distribution functions of other quantities,
such as relaxation time, local and global density of state. It is
found that the structure of an ALS may be different, depending on the
specific quantity, for which it constitutes an optimal
fluctuation. Relations between various procedures of selection of ALS,
and between asymptotics of corresponding distribution functions, are
discussed.

\end{abstract}
\narrowtext

\section{Introduction}
\label{s1}

Mesoscopic fluctuations of various physical quantities in disordered
systems have been intensively investigated during the last decade
\cite{ALW}. These fluctuations originating from the quantum coherence
of wave functions are typically much stronger than what usual
statistical considerations would predict. In particular, it was found
by Altshuler, Kravtsov and Lerner \cite{akl} that distribution
functions of conductance, density of states (DOS), local density of
states (LDOS), and relaxation times have slowly decaying
logarythmically--normal (LN) asymptotics at large values of the
arguments. These results were obtained within the renormalization
group (RG) treatment of the field--theoretical $\sigma$--model
describing the low--momenta physics of the problem. The validity of
this RG approach is restricted to 2D and $2+\epsilon$--dimensional
systems, with $\epsilon\ll 1$.

On the other hand, the conductance, LDOS and relaxation times
fluctuations in strictly 1D disordered chains, where all states are
strongly localized, were studied with the use of Berezinski and
Abrikosov--Ryzhkin techniques \cite{abr,alpr,alpr2}. The corresponding
distributions were found to be of the LN form, too. It was conjectured on
the basis of this similarity \cite{akl,alpr,krav}
that even in a metallic sample there is a
finite probability to find ``almost localized'' eigenstates.
Precise meaning of these words remained however obscure.

More recently, the interest to statistical properties of
eigenfunctions in disordered and chaotic systems started to grow.
On the experimental side, it was motivated by the possibilty of
fabrication of small systems (quantum dots)
with well resolved electron energy levels \cite{dots}.
Conductance fluctuations of such a dot are related to statistical
properties of wave function amplitudes \cite{jal}. Besides, the
microwave cavity technique \cite{microwave} allows to observe
experimentally spatial fluctuations of electromagnetic wave amplitude
in chaotis or disordered cavities \cite{kud}.

On the theoretical side, the recent progress is due to the observation
\cite{mf} that the statistics of eigenfunctions intensities can be
very efficiently studied with making use of the supersymmetry
technique. This allows one to reformulate the problem in terms of the
supermatrix $\sigma$--model \cite{note1}.
It was found that the zero--dimensional
approximation to this $\sigma$--model reproduces the Random Matrix
Theory (RMT) results. In the case of quasi--1D geometry, the model was
solved exactly by means of the transfer--matrix method \cite{mf2},
which allowed us to calculate various statistical properties of
eigenfunctions \cite{mf,mf2,mf3,FMRev,mf4}. It was found that the
distribution function ${\cal P}(u)$ of eigenfunction intensities
$u=|\psi^2(\bbox{r})|$ deviates slightly from its RMT form for not
too large $u$, but decreases much  slower  than RMT predicts, in its
far ``tail''. In $d>1$ dimensions, the results for ${\cal P}(u)$ were
obtained by means of perturbative method \cite{mf4} and saddle--point
approximation \cite{falef}, and are qualitatively similar to the
quasi--1D behavior.

It should be noted that the saddle--point method for the supermatrix
$\sigma$--model used in Ref.\cite{falef} and suggested previously
by Muzykantskii and Khmelnitskii \cite{mk} turned out to be a powerful
tool to study the asymptotic ``tails'' of various distributions. In
particular, it was used in \cite{mk,relax} to study the long--time
relaxation phenomena, and in \cite{ldos-as} to find the asymptotical
behavior of the distribution of LDOS. The obtained decrease rates are
much slower than those given by the perturbation theory. In 2D, the
far asymptotics are of LN type \cite{relax,ldos-as}, in agreement with
the RG results of \cite{akl}. The asymptotic behavior of the
distributions was again attributed to the effect of almost localized
states \cite{mk,mf4,falef}. Moreover, it was conjectured in
\cite{mk,falef} that the form of the saddle-point solution of the
$\sigma$--model directly describes the electron density of such a
state.

The purpose of this paper is to study in most detail the spatial
structure of the anomalously localized states (ALS). The paper is
organized as follows. In section \ref{s2} we review briefly the previously
obtained results for the distribution of eigenfunctions intensity.
In section \ref{s3}, which is the central one for the paper, we study the
case of a quasi--1D geometry of the sample. We present the exact
calculation of the
properties of ALS responsible for the asymptotics of ${\cal P}(u)$. In
subsection \ref{s3.1.1} the average intensity
$\langle|\psi^2(\bbox{r})|\rangle$ of such a state is found for a
sample with the length $L\gg\xi$, where $\xi$ is the localization
length. The state with an anomalously high local intensity
$|\psi^2(0)|=u$ is characterized by an effective localization
length $\xi_{ef}\sim(\xi/uS)^{1/2}$ (here $S$ is the wire cross-section),
where most of its normalization is concentrated. Outside this central
region, $\langle|\psi^2(r)|\rangle$ decreases as $1/r^2$.
Finally, in the vicinity
of the observation point there is a sharp change of
$\langle|\psi^2|\rangle$ from $|\psi^2(0)|=u$ to
$\langle|\psi^2|\rangle=(u/\xi S)^{1/2}/2$. In subsection \ref{s3.1.2} we
repeat this calculation in the case of metallic sample with $L\ll
\xi$. We find that ALS have exactly the same form of
$\langle|\psi^2(r)|\rangle$, if $u\gg\xi/SL^2$. The latter condition
means that $\xi_{ef}\ll L$, i.e. the state plays a role of a localized
one even from the point of view of a short sample with the length
$L$.

In subsection \ref{s3.2} we study the fluctuations of intensity
$|\psi^2(r)|$ of an ALS. We find that these fluctuations are of the
RMT form. The narrow region around the observation point where the
quasi-jump of  $\langle|\psi^2|\rangle$
occurs, is an exception; here the fluctuations
are suppressed.

Comparing our results with the corresponding saddle-point solution
\cite{falef}, we conclude that the latter describes exactly the form
of the average density $\langle|\psi^2(r)|\rangle$, up to a
normalization factor and without the quasi-jump around the observation
point $r=0$. This allows us to generalize the results to the spatial
dimension $d>1$ (section \ref{s4}). In section \ref{s5} we discuss and compare
the spatial structure of ALS corresponding to asymptotical behavior of
various  distribution functions. In Sec.\ref{s6} we discuss briefly
some special features of the ALS located close to the sample edge, and
dependence of the results on the symmetry of the ensemble.
Our results are summarized in section
\ref{s7}.

\section{Distribution of eigenfunction amplitudes: Overview of
results.}
\label{s2}

In this section, we have collected some results for the eigenfunction
intensity distribution function obtained with use of the supermatrix
$\sigma$--model formalism. The $\sigma$--model action reads
\begin{equation}
S[Q]=-{\beta\over 2}
\int d^dr\,\mbox{Str}\left[{\pi\nu_{0} D\over 4}(\nabla
Q)^2-\pi\nu_{0}\eta\Lambda Q\right]\ .
\label{2}
\end{equation}
Here $Q(\bbox{r})$ is a $4\times 4$ or $8\times8$ supermatrix field
for the cases of
unitary symmetry (broken time reversal invariance) and orthogonal
symmetry (unbroken time reversal invariance) respectively. We will
label the formulas for the first case by the index U, and for the
second case by the index O. The matrix $\Lambda$ is defined as
$\Lambda=\mbox{diag}\{1,1,-1,-1\}$ (U) or
$\Lambda=\mbox{diag}\{1,1,1,1,-1,-1,-1,-1\}$ (O) and the coefficient $\beta$
is equal to $\beta=2$ (U) or $\beta=1$ (O). Furthermore,
$\mbox{Str}$ stands for the supertrace (trace over bosonic components minus
 trace over fermionic ones), $D$ is the diffusion constant,
$\eta$ is the level broadening
(or, else, $(-2i)\times$ frequency) and $\nu_{0}$ is the mean DOS.
We do not go into details of the supersymmetric formalism here, which
can be found e.g. in \cite{efe,vwz,FMRev}.
Let us define now
the function $Y(Q_0)$ as
\begin{equation}
Y(Q_0)=\int_{Q(\bbox{r_0})=Q_0} DQ(r)
\exp\{-S[Q]\}\ .
\label{1}
\end{equation}
For the invariance reasons, the function $Y(Q_0)$ turns out to be
dependent in the unitary symmetry case on the two scalar variables
$1\le \lambda_{1}<\infty $ and $ -1\le \lambda_{2} \le 1$ only, which are
eigenvalues of the ``retarded-retarded'' block of the matrix $Q_0$.
Moreover, in the limit $\eta \to 0$ (at a fixed  value of the
system volume) only the dependence on $\lambda_1$ persists:
\begin{equation}
Y(Q_0)\equiv Y(\lambda_1,\lambda_2)\to Y_a(2\pi\nu_{0} \eta\lambda_1)
\label{3}
\end{equation}
The distribution ${\cal P}(u)$ of the eigenfunctions intensities
$u=|\psi(\bbox{r_0}|^2$ is then given by
\cite{mf,mf5}
\begin{equation}
{\cal P}(u)={1\over V}{d^2 \over du^2} Y_a(u)
      ={1\over V} {d^2 \over du^2} Y\left.(\lambda_1={u\over
2\pi\nu_{0} \eta})\right|_{\eta\to 0} \qquad (U)
\label{4}
\end{equation}
In the case of orthogonal symmetry, $Y(Q_0)\equiv
Y(\lambda_1,\lambda_2,\lambda)$, where
$1\le \lambda_{1},\lambda_{2} <\infty $ and $ -1\le \lambda \le 1$.
In the limit $\eta\to 0$, the relevant region of values is
$\lambda_1\gg\lambda_2,\lambda$, where
\begin{equation}
Y(Q_0)\to Y_a(\pi\nu_{0}\eta\lambda_1)
\label{3o}
\end{equation}
The distribution of eigenfunction intensities is in this case
expressed through the function $Y_a$ as follows \cite{mf}:
\begin{eqnarray}
{\cal P}(u)&=& {1\over\pi Vu^{1/2}}\int_{u/2}^{\infty}
dz (2z-u)^{-1/2}{d^2\over dz^2} Y_a(z) \nonumber\\
&=& {2\sqrt{2}\over\pi Vu^{1/2}}{d^2\over du^2}\int_0^\infty {dz\over
z^{1/2}}  Y_a(z+u/2) \qquad (O) \label{4o}
\end{eqnarray}

In the diffusive sample, typical configurations of the $Q$--field are
nearly constant in space, so that one can approximate the functional
integral  (\ref{1}) by an integral over a single supermatrix $Q$. This
procedure, which makes the problem effectively zero-dimensional and is
known as zero-mode approximation, gives
\begin{equation}
Y_a(z)\approx e^{-Vz} \qquad (O,U)\ ,
\label{anom1}
\end{equation}
and consequently,
\begin{eqnarray}
&{\cal P}(u)\approx Ve^{-uV}&\qquad (U)\ , \label{anom2u}\\
&{\cal P}(u)\approx \sqrt{{V\over 2\pi u}}e^{-uV/2} & \qquad (O)\
, \label{anom2o}
\end{eqnarray}
which are just the RMT results for the Gaussian Unitary Ensemble (GUE)
and Gaussian Orthogonal Ensemble (GOE) respectively \cite{mehta}.

For the case of quasi--1D geometry one can solve the problem exactly,
by evaluating the integral in (\ref{1}) with making use of the
transfer-matrix method \cite{mf}. As a result, one gets an expression
for ${\cal P}(u)$ depending on a scaling parameter $x=L/\xi$, where $L$
is the sample length, $\xi=2\pi\nu_{0} SD$ is the localization length,
and $S$ is the transverse cross-section \cite{mf,FMRev,mf4}.
In particular,
in the case of a short (metallic) sample, $x\ll 1$, we find
 for the distribution of normalized intensity
$y=uV$:
\begin{eqnarray}
&{\cal P}^{(U)}(y)= e^{-y}\left[1+{\alpha x\over 6}(2-4y+y^2)+
\ldots\right]\
;\qquad & y\lesssim x^{-1/2} \label{adm15a}\\
&{\cal P}^{(O)}(y)=\frac{e^{-y/2}}{\sqrt{2\pi y}}
\left[1+{\alpha x\over 6}\left({3\over 2}-3y+{y^2\over 2}\right)
+\ldots\right]\ ;\qquad & y\lesssim x^{-1/2} \label{adm8a} \\
&{\cal P}(y)\sim \exp\left\{ {\beta\over 2}
\left [-y+{\alpha \over 6}y^2x+\ldots\right]\right\}\
;\qquad & x^{-1/2}\lesssim y\lesssim x^{-1} \label{adm15b}\\
&{\cal P}(y)\sim\exp\left[-2\beta\sqrt{y/x}\right]\ ;\qquad & y\gtrsim
x^{-1} \label{adm15c}
\end{eqnarray}
Here the coeficient $\alpha$ is equal to
$\alpha=2[1-3L_-L_+/L^2]$, where $L$ is the sample length and
$L_-$, $L_+$ are the distances from the observation point $\bbox{r_0}$ to
the sample edges. If one is interested in the distribution function
${\cal P}(y)$ averaged over the position of the point $\bbox{r_0}$
within the sample, one has to replace  $\alpha$ in eqs.(\ref{adm15a}),
(\ref{adm8a}) by its average value equal to unity, and in
eq.(\ref{adm15b}) by its maximum value equal to $2$. Such averaged
distribution functions were considered in our previous publications,
Ref.\cite{mf,FMRev,mf4}. It is straightforward, however, to generalize
the derivation and to get the formulas (\ref{adm15a})--(\ref{adm15b})
for the position-dependent distribution functions.
 Note that $g=1/x$ is for small $x$ just the dimensionless conductance of
the sample: $g=2\pi\nu_{0} DS/L = G/(e^2/h)$. We see, that for not too
large amplitudes $y$, Eqs.(\ref{adm15a}), (\ref{adm8a}) are valid
which are just the RMT results with relatively small corrections.
In the intermediate region (\ref{adm15b}), the correction {\it in the
exponent} is small compared to the leading term but much larger than
unity, so that ${\cal P}(y)\gg {\cal P}_{RMT}(y)$ though
$\ln{\cal P}(y)\simeq \ln{\cal P}_{RMT}(y)$. Finally, in the large
amplitude region,
(\ref{adm15c}), the distribution function ${\cal P}(y)$ differs
completely from the RMT prediction. We will present eq.(\ref{adm15c})
in a more precise form in Sec.\ref{s3}. Note that it is not valid when
the observation point is located close to the sample boundary, as will
be explained in Sec.\ref{s6.1}.
The asymptotic behavior (\ref{adm15c}) is the same as in the
case of insulating sample, $L\gg\xi$, when the distribution
${\cal P}(u)$ takes the form
\begin{eqnarray}
&&{\cal P}^{(U)}(u)\simeq {8\xi^2 S\over L} \left[K_1^2(2\sqrt{uS\xi})+
K_0^2(2\sqrt{uS\xi})\right]  \label{anom3u}\\
&& {\cal P}^{(O)}(u)\simeq {2\xi^2 S\over L} {K_1(2\sqrt{uS\xi})\over
\sqrt{uS\xi} } \label{anom3o}
\end{eqnarray}
This allowed us to conjecture in Ref.\cite{mf4} that the asymptotic
behavior
(\ref{adm15c}) is controlled by the probability to have a
quasilocalized eigenstate with an effective spatial extent much less
than $\xi$. We will prove this rigorously in the present paper.

In $d>1$ dimensions the problem can not be solved exactly anymore, and
some approximate methods of evaluation of the integral (\ref{1}) are
necessary. For moderately large amplitudes $y\lesssim \kappa^{-1/2}$,
a perturbative treatment of  non-zero modes of the
$\sigma$--model in the weak localization region
is possible \cite{mf4}. Here $\kappa$ is the usual
parameter of the perturbation theory \cite{akl} defined as
\begin{equation}
\kappa=\sum_{\bbox{q}}{1\over 2\pi\nu_{0} D \bbox{q}^2V}\ ;
\label{anom4}
\end{equation}
the summation goes over the eigenmodes of the Laplace operator in the
sample. The result for ${\cal P}(y)$  is:
\begin{eqnarray}
&&{\cal P}^{(U)}(y)=e^{-y}[1+\kappa(2-4y+y^2)+\ldots]
\label{anom5u}\\
&& {\cal P}^{(O)}(y)=\frac{e^{-y/2}}{\sqrt{2\pi y}}
\left[1+\kappa\left({3\over 2}-3y+{y^2\over 2}\right)
+\ldots\right]\label{anom5o}
\end{eqnarray}
In particular, in quasi--1D systems $\kappa=1/6g$, and
Eqs.(\ref{adm15a}), (\ref{adm8a}) are reproduced.
For $d \ge 2$ the corresponding  sum over momenta $\bbox{q}$ diverges
at large $|\bbox{q}|$ and is to be cut off at $|\bbox{q}|\sim l^{-1}$.
This gives $\kappa=\frac{1}{4\pi^2\nu_{0} D}\ln(L/l)$ for $d=2$ and
$\kappa\sim(k_F l)^2$ for $d=3$, where $k_F$ is the Fermi momentum.

In the region of large amplitudes $y\gtrsim\kappa^{-1/2}$,
Eq.(\ref{anom5u}), (\ref{anom5o}) loose their validity. In this
region, the distribution ${\cal P}(y)$ can be found \cite{falef} by
using the saddle-point approximation. For $d=2,3$ the result is
\begin{eqnarray}
& {\cal P}(y)\sim\exp\left\{ {\beta\over 2} (-y+\kappa
y^2+\ldots)\right\}\ ,\qquad &\kappa^{-1/2}\lesssim y\lesssim
\kappa^{-1}
\label{anom6}\\
&{\cal P}(y)\sim\exp\left\{- {\beta\over 8\kappa} \ln^d(\kappa y)\right\}\
, \qquad & y\gtrsim\kappa^{-1} \label{anom7}
\end{eqnarray}

In the next two sections, we will study in detail the structure of the states,
which are responsible for the far asymptotics (\ref{adm15c}), (\ref{anom7}).
For definitness, we will consider the unitary symmetry case;
generalization to the orthogonal ensemble is completely
straightforward, and we will just quote the (minor) changes in the
results in Sec.\ref{s6.2}.

\section{Anomalously localized states in quasi--1D samples}
\label{s3}
\setcounter{equation}{0}
\subsection{Average form of the eigenfunction corresponding to an
anomalously high local amplitude.}
\label{s3.1}

In this section, we study the average intensity
$\langle|\psi^2(\bbox{r})|\rangle_u$ of a wave function with the
condition that $|\psi^2(0)|=u$. For convenience, we have put the
observation point in the coordinate origin; the sample edges are at
$r=-L_-$ and $r=L_+$; the sample length is $L=L_-+L_+$. Formally,
$\langle|\psi^2(\bbox{r})|\rangle_u$ is defined as
\begin{equation}
\langle|\psi^2(\bbox{r})|\rangle_u = \frac {{\cal Q}(u,\bbox{r})}
{{\cal P}(u)}\ ,
\label{anom8}
\end{equation}
where
\begin{eqnarray}
{\cal P}(u)&=& {1\over\nu_{0}
V}\langle\sum_\alpha
\delta(|\psi_\alpha(0)|^2-u)\delta(E-E_\alpha)\rangle
\label{anom9}\\
{\cal Q}(u,\bbox{r})&=& {1\over\nu_{0}
V}\langle\sum_\alpha |\psi_\alpha(\bbox{r})|^2
\delta(|\psi_\alpha(0)|^2-u)\delta(E-E_\alpha)\rangle\ ,
\label{anom10}
\end{eqnarray}
where $\psi_\alpha$ are  eigenfunctions of the Hamiltonian, and
$E_\alpha$ are the corresponding eigenvalues.
The here defined function ${\cal P}(u)$ is just the distribution
function of local intensity discussed in the preceding section.
As has been mentioned above, it was calculated in the quasi--1D case,
with making use of the $\sigma$--model representation, eqs.(\ref{4}),
(\ref{4o}), and the transfer-matrix method \cite{mf,FMRev}.
The result reads
\begin{equation}
{\cal P}(u)={1\over V} {\partial^2\over \partial u^2}\left\{
W^{(1)}(uS\xi,\tau_+)W^{(1)}(uS\xi,\tau_-)\right\}\ ,
\label{anom11}
\end{equation}
where the function $W^{(1)}(z,\tau)$ satisfies the equation
\begin{equation}
{\partial W^{(1)}(z,\tau)\over \partial\tau}=
\left( z^2 {\partial^2\over \partial z^2}-z\right)
W^{(1)}(z,\tau)
\label{anom12}
\end{equation}
and the boundary condition
\begin{equation}
W^{(1)}(z,0)=1
\label{anom13}
\end{equation}
The solution to eqs.(\ref{anom12}), (\ref{anom13}) can be found in
terms of the Lebedev--Kontorovich expansion
\begin{equation}
W^{(1)}(z,\tau)=2z^{1/2}\left\{K_{1}(2z^{1/2})+\frac{2}{\pi}\int_
{0}^{\infty}d\nu\frac{\nu}{1+\nu^{2}}
\sinh{\frac{\pi\nu}{2}}K_{i\nu}(2z^{1/2})
e^{-\frac{1+\nu^{2}}{4}\tau}\right\}
\label{anom14}
\end{equation}
Now we turn to evaluation of the function ${\cal Q}(u,r)$
defined in eq.(\ref{anom10}). Detailed exposition of the method used
can be found in Ref.\cite{FMRev}, and we will mainly
follow the notations of this paper. We start from expressing the
moments
$\langle |\psi(\bbox{r})|^2 |\psi(0)|^{2q}\rangle$ in terms of the
Green's functions
\begin{eqnarray}
&&\langle |\psi(\bbox{r})|^2 |\psi(0)|^{2q}\rangle \nonumber\\
&&\equiv {1\over \nu_{0} V}
\langle\sum_\alpha |\psi_\alpha(\bbox{r})|^2 |\psi_\alpha(0)|^{2q}
\delta(E-E_\alpha)\rangle \nonumber \\
&&=\lim_{\eta\to 0} {i^{q-1}\over 2\pi\nu_{0} V} (2\eta)^q \langle
G_R^q(0,0) G_A(\bbox{r},\bbox{r})\rangle\ ,
\label{anom15}
\end{eqnarray}
where
\begin{equation}
G_{R,A}(\bbox{r_1},\bbox{r_2})=\langle
\bbox{r_1}|(E-\hat{H}\pm i\eta)^{-1}|\bbox{r_2}\rangle
\label{anom16}
\end{equation}
Here $\hat{H}$ is the Hamiltonian which consists of the free part
$\hat{H}_0$ and the disorder potential $U(\bbox{r})$:
\begin{equation}
\hat{H}=\hat{H}_0+U(\bbox{r})\ ;\qquad \hat{H}_0={1\over 2m}
\hat{\bbox{p}}^2\ ; \label{anom18}
\end{equation}
the latter being defined by the correlator
\begin{equation}
\langle U(\bbox{r})U(\bbox{r'})\rangle={1\over
2\pi\nu_{0}\tau_s}\delta(\bbox{r}-\bbox{r'})
\label{anom19}
\end{equation}
Next, we write the  product of the Green's functions in terms
of the integral over a supervector field
$\Phi=(S_1,S_2,\chi_1,\chi_2)$:
\begin{eqnarray}
G_R^q(0,0) G_A(\bbox{r},\bbox{r})&=& {i^{1-q}\over q!}\int D\Phi\,
D\Phi^\dagger (S_1(0) S_1^*(0))^q S_2(\bbox{r})S_2^*(\bbox{r})
\nonumber\\&\times &
\exp\left\{i\int d\bbox{r'} \Phi^\dagger(\bbox{r'}) \Lambda^{1/2}
(E+i\eta\Lambda-\hat{H})\Lambda^{1/2} \Phi(\bbox{r'})\right\}
\label{anom17}
\end{eqnarray}
The following steps are \cite{FMRev}:
\begin{itemize}
\item[i)] averaging over the disorder;
\item[ii)] introducing a $4\times 4$ supermatrix variable
$R_{\mu\nu}(\bbox{r})$ having the symmetry of the tensor product
$\Phi_\mu(\bbox{r}) \Phi^\dagger_\nu(\bbox{r})$ ;
\item[iii)] integrating out the $\Phi$ fields ;
\item[iy)] using the saddle-point approximation which leads to the
following equation for $R$:
\begin{eqnarray}
&& R(\bbox{r})={1\over 2\pi\nu_{0}\tau_s} g(\bbox{r},\bbox{r})\ ;
\label{anom20}\\
&&g(\bbox{r_1},\bbox{r_2})=\langle
\bbox{r_1}|(E-\hat{H}_0-R)^{-1}|\bbox{r_2}\rangle
\label{anom21}
\end{eqnarray}
\end{itemize}
The relevant set of the solutions (the saddle-point manifold)  has the
form:
\begin{equation}
R=\sigma\cdot I - {i\over 2\tau_s} Q
\label{anom22}
\end{equation}
where $I$ is the unity matrix, $\sigma$ is certain constant,
and $Q$ belongs to the coset space
$U(1,1|2)$ and satisfies the condition $Q^2=1$. Finally, we find
\begin{eqnarray}
\langle |\psi(\bbox{r})|^2 |\psi(0)|^{2q}\rangle&=&-{1\over
2V}\lim_{\eta\to 0} (2\pi\nu_{0}\eta)^q\int DQ
\left[Q_{11,bb}^q(0)Q_{22,bb}(\bbox{r})\right. \nonumber\\
&+& \left. q{1\over(\pi\nu_{0})^2} Q_{11,bb}^{q-1}(0)
g_{12,bb}(0,\bbox{r}) g_{21,bb}(\bbox{r},0)\right]e^{-S\{Q\}}\ ,
\label{anom22a}
\end{eqnarray}
where $S[Q]$ is the $\sigma$--model action defined in eq.(\ref{1}).

Taking into account that $Q(\bbox{r})$ varies slowly on a scale of the
mean free path $l$, we have for $|\bbox{r_1}-\bbox{r_2}|\ll l$
\begin{equation}
g(\bbox{r_1},\bbox{r_2})\simeq \mbox{Re}
\langle G_R(\bbox{r_1}-\bbox{r_2})\rangle
+i Q(\bbox{r_1})\mbox{Im}
\langle G_R(\bbox{r_1}-\bbox{r_2})\rangle
\label{anom23}
\end{equation}
where
\begin{equation}
\langle G_R(\bbox{r_1}-\bbox{r_2})\rangle=
\langle \bbox{r_1}|(E-\hat{H}_0-\rho+{i\over
2\tau_s})^{-1}|\bbox{r_2}\rangle
\label{anom24}
\end{equation}
is the average one-particle Green's function. In the opposite limiting
case $|\bbox{r_1}-\bbox{r_2}|\gg l$, the Green's function
$g(\bbox{r_1},\bbox{r_2})$ vanishes exponentially. Thus, we get
\begin{equation}
\langle |\psi(\bbox{r})|^2 |\psi(0)|^{2q}\rangle \simeq -{1\over
2V}\lim(2\pi\nu_{0}\eta)^q\int DQ\,
Q^q_{11,bb}(0)Q_{22,bb}(\bbox{r})e^{-S\{Q\}}\
,\qquad r\gg l
\label{anom25}
\end{equation}
\begin{eqnarray}
\langle |\psi(\bbox{r})|^2 |\psi(0)|^{2q}\rangle &\simeq& -{1\over
2V}\lim(2\pi\nu_{0}\eta)^q\int DQ\,
\left[Q^q_{11,bb}(0)Q_{22,bb}(0) \right.\nonumber\\
&+& \left. q k_d(r)Q_{11,bb}^{q-1}(0)Q_{12,bb}(0)Q_{21,bb}(0)\right]
e^{-S\{Q\}}\ ,\qquad r\ll l
\label{anom26}
 \end{eqnarray}
where
\begin{equation}
k_d(r)= {|\mbox{Im}\langle G_R(r)\rangle|^2\over (\pi\nu_{0})^2}
\label{anom27}
\end{equation}
depends on short-scale dimensionality $d$ of the sample \cite{note2}.
In particular, for a strip ($d=2$) or a wire ($d=3$) we have
\begin{eqnarray}
&& k_2(r)=J_0^2(k_F r)\ , \label{anom28}\\
&& k_3(r)={\sin^2(k_F r)\over (k_F r)^2}\ , \label{anom29}
\end{eqnarray}
respectively.

Now we evaluate the $Q$--integrals in eqs.(\ref{anom25}),
(\ref{anom26}) using the transfer-matrix method \cite{FMRev}. We first
consider the $r\gg l$ case. We will assume that the transverse size of
the wire is much less than the effective localization length of the
ALS, which we will find to be $\xi_{ef}\sim\sqrt{\xi/ uS}$. This
means that we are indeed in the quasi--1D regime, when considering the
ALS structure.  We assume for definitness
that the point r is located to the positive direction from the
coordinate origin: $-L_-<0<r<L_+$. We get then:
\begin{equation}
\langle |\psi(\bbox{r})|^2 |\psi(0)|^{2q}\rangle=
{1\over V(S\xi)^q} q\int_0^\infty dz\,z^{q-2} W^{(1)}(z,\tau_-)
W^{(2)}(z,\tau_1,\tau_2)\ ,
\label{anom30}
\end{equation}
where $\tau_-=L_-/\xi$, $\tau_1=r/\xi$, $\tau_2=(L_+-r)/\xi$; the
function $W^{(1)}(z,\tau)$ is defined by
eqs.(\ref{anom12})--(\ref{anom14}),
and the function  $W^{(2)}(z,\tau_1,\tau_2)$ satisfies the same equation
\begin{equation}
{\partial W^{(2)}(z,\tau_1,\tau_2)\over \partial\tau_1}=
\left( z^2 {\partial^2\over \partial z^2}-z\right)
W^{(2)}(z,\tau_1,\tau_2)
\label{anom31}
\end{equation}
and the boundary condition
\begin{equation}
W^{(2)}(z,0,\tau_2)=zW^{(1)}(z,\tau_2)
\label{anom32}
\end{equation}
The solution of eqs.(\ref{anom31}), (\ref{anom32})
is \cite{FMRev}
\begin{eqnarray}
&&W^{(2)}(z,\tau_1,\tau_2)=2\sqrt{z}\int_0^\infty d\nu
b(\nu,\tau_2)K_{i\nu}(2\sqrt{z}) e^{-{1+\nu^2\over 4}\tau_1}\ ;
\nonumber\\
&& b(\nu,\tau_2)={\nu\sinh(\pi\nu)\over 2 \pi^2}\int_0^\infty dt\,
K_{i\nu}(t) W^{(1)}(t^2/4,\tau_2)
\label{anom33}
\end{eqnarray}
Substituting here the formula (\ref{anom14}) for $W^{(1)}(z,\tau_2)$
and evaluating the integral over $z$, we reduce eq.(\ref{anom33}) for
$b(\nu,\tau_2)$ to the form
\begin{eqnarray}
b(\nu,\tau_2)&=&{\nu\sinh(\pi\nu)\over 16\pi^2}
\left|\Gamma\left({1+i\nu\over 2}\right)\right|^4(1+\nu^2)\nonumber\\
&+&{\nu\sinh(\pi\nu)\over 2\pi^3}\int {d\nu_1\,\nu_1\over 1+\nu_1^2}
\sinh{\pi\nu_1\over 2}
\left|\Gamma\left(1+i{\nu+\nu_1\over 2}\right)\right|^2
\left|\Gamma\left(1+i{\nu-\nu_1\over 2}\right)\right|^2
e^{-{1+\nu_1^2\over 4}\tau_2}
\label{anom34}
\end{eqnarray}
Equations (\ref{anom30}), (\ref{anom14}), (\ref{anom33}),
(\ref{anom34}) constitute the final result for the moments
$\langle |\psi(\bbox{r})|^2 |\psi(0)|^{2q}\rangle$ at $r\gg l$.
Now we can restore the function ${\cal Q}(u,\bbox{r})$:
\begin{eqnarray}
{\cal Q}(u,\bbox{r})&\equiv &
\langle\delta(|\psi(0)|^2-u)|\psi(\bbox{r})|^2\rangle \nonumber\\
& =& -{1\over V\xi S}{\partial\over\partial u}
\left[{W^{(2)}(u\xi S,\tau_1,\tau_2)W^{(1)}(u\xi S,\tau_-)\over
u}\right]\ ,\qquad r\gg l
\label{anom35}
\end{eqnarray}

In the opposite case, $r\ll l$, we find from eq.(\ref{anom26})
\begin{equation}
{\cal Q}(u,\bbox{r})={1\over V}\left\{k_d(r)\left(u{d^2\over du^2}+
{d\over du}\right)-{d\over du}\right\} Y_a(u)
\ , \label{anom36}
\end{equation}
where the function $Y_a(u)$ was defined in eq.(\ref{3}). This formula
is valid for any sample, which is locally $d$--dimensional. In the
case of the quasi--1D geometry we get
\begin{equation}
{\cal Q}(u,\bbox{r})={1\over V}\left\{k_d(r)\left(u{d^2\over du^2}+
{d\over du}\right)-{d\over du}\right\}
\left[W^{(1)}(u\xi S,\tau_-)W^{(1)}(u\xi S,\tau_+)\right]\ ,\qquad
r\ll l
\label{anom37}
\end{equation}

Substituting these results along with the formula (\ref{anom11}) for
${\cal P}(u)$ in eq.(\ref{anom8}), we can find the average density
$\langle |\psi^2(\bbox{r})|\rangle_u$. It is possible to check that it
satisfies the normalization requirment
$\int \langle |\psi^2(\bbox{r})|\rangle_u = 1$, as it should be.
In the following two subsections, we will study the form of
$\langle |\psi^2(\bbox{r})|\rangle_u$ for an ALS in the insulating
($L\gg\xi$) and the metallic ($L\ll\xi$) limits.

\subsubsection{Insulating sample ($L\gg\xi$).}
\label{s3.1.1}
In this subsection, we analyze the above general results for
$\langle |\psi^2(\bbox{r})|\rangle_u$ in the limit of an infinite
sample length. More precisely, we will assume that the distances
$L_-,L_+$ from the
point $\bbox{r}=0$ to the both edges of the sample are much larger than
$\xi$, so that we can consider the limit $\tau_-,\tau_+\to\infty$. We have
then
\begin{equation}
W^{(1)}(z,\infty)=2\sqrt{z}K_1(2\sqrt{z})
\label{anom38}
\end{equation}
and the distribution function ${\cal P}(u)$ takes the form
(\ref{anom3u}).
Furthermore, eqs.(\ref{anom33}), (\ref{anom34}) reduce to
\begin{equation}
W^{(2)}(z,\tau_1,\infty)=2\sqrt{z}\int_0^\infty d\nu{\nu(1+\nu^2)\over
8}\tanh{\pi\nu\over 2} K_{i\nu}(2\sqrt{z})e^{-{1+\nu^2\over 4}\tau_1}
\label{anom39}
\end{equation}
Typical values for the intensity $u$ of a localized state are
$u\lesssim1/S\xi$. We are interested however in anomalously high $u\gg
1/S\xi$. It is seen from eqs.(\ref{anom11}), (\ref{anom14}),
(\ref{anom35}), (\ref{anom39}), that this corresponds to large values
$t=2\sqrt{uS\xi}\gg 1$ of the argument  of the modified Bessel functions
$K_1(t)$ and $K_{i\nu}(t)$. The corresponding asymptotic formulae read
\cite{bateman}:
\begin{eqnarray}
&&K_1(t)\simeq\sqrt{{\pi\over 2t}}e^{-t}\ ,\qquad t\gg 1
\label{anom40}\\
&&K_{i\nu}(t)\simeq \sqrt{{\pi\over 2t}}\exp
\left\{-t-tf\left({\nu\over t}\right)\right\}\ ,\qquad 1\ll t\ ,\ \
\nu<t
\label{anom41}\\
&&f(w)=\sqrt{1-w^2}+w\arcsin{w}-1={w^2\over 2} + O(w^4)
\nonumber
\end{eqnarray}
The asymptotics of the distribution function ${\cal P}(u)$ has the form
\begin{equation}
V{\cal P}(u)\simeq 4\pi S^{3/2}\xi^{3/2}u^{-1/2}e^{-4\sqrt{uS\xi}}
\label{anom42}
\end{equation}
As to the function ${\cal Q}(u,r)$, we consider separately three regions
of the distance $r$:

a) $r\gg\xi$, i.e. $\tau_1=r/\xi\gg 1$. \\
The integral in (\ref{anom39}) is then determined by the region
$\nu\sim\tau_1^{-1/2}$, yielding
\begin{equation}
W^{(2)}(z=t^2/4,\tau_1,\infty)\simeq{\pi^2\over 8\sqrt{2}}
t^{1/2}\tau_1^{-3/2}e^{-t-\tau_1/4}\ ,
\label{anom43}
\end{equation}
where we have denoted $t=2\sqrt{z}$. Consequently,
\begin{equation}
V{\cal Q}(u,r)={\pi^{5/2}\over 4u}\tau_1^{-3/2}e^{-4\sqrt{uS\xi}-\tau_1/4}
\label{anom44}
\end{equation}

b) $l\ll r\ll \xi$.
\\We have
\begin{eqnarray}
W^{(2)}(z=t^2/4,\tau_1,\infty)
&\simeq&{1\over 8}\sqrt{\pi t\over 2}e^{-t-\tau_1/4}
\int_0^\infty d\nu\,\nu(1+\nu^2)\tanh{\pi\nu\over
2}\exp\left\{-\nu^2\left({1\over 2t}+{\tau_1\over 4}\right)\right\}
\nonumber\\
&=&\sqrt{{\pi\over 2}}t^{5/2}{1\over(\tau_1 t+2)^2}e^{-t}\ ,
\label{anom45}
\end{eqnarray}
so that
\begin{equation}
V{\cal Q}(u,r)=2\pi\xi S e^{-4\sqrt{uS\xi}} {1\over
(1+\tau_1\sqrt{uS\xi})^2}
\label{anom46}
\end{equation}

c) $r\ll l$.  \\
In this region an additional term proportional to the Friedel function
$k_d(r)$ appears in the expression for ${\cal Q}(u,r)$, see
eq.(\ref{anom37}). We get
\begin{equation}
V{\cal Q}(u,r)=2\pi\xi S e^{-4\sqrt{uS\xi}}
\left[1+2\sqrt{uS\xi} k_d(r)\right]
\label{anom47}
\end{equation}
Since $k_d(r)\sim 1/(k_F r)^{d-1}$ for $r\gg k_F^{-1}$, the second
term in square brackets is much larger than the first one at
$r\ll r_0\sim k_F^{-1}(uS\xi)^{1/[2(d-1)]}$ and is negligible at $r\gg
r_0$.

Substituting now eqs.(\ref{anom42}), (\ref{anom44}), (\ref{anom46}),
and (\ref{anom47}) in eq. (\ref{anom8}), we find the following spatial
structure of the ALS with $|\psi^2(0)|=u$:
\begin{eqnarray}
&\langle|\psi^2(r)|\rangle_u=
{\pi^{3/2}\over 16}u^{-1/2}S^{-3/2}r^{-3/2}e^{-r/4\xi}\ ,&\qquad
r\gg\xi \label{anom48a}\\
&\langle|\psi^2(r)|\rangle_u=
\displaystyle{
{1\over 2}\left({u\over\xi S}\right)^{1/2} {1\over
\left(1+r\sqrt{uS\over\xi}\right)^2 }    }
\ ,&\qquad l<r\ll\xi
\label{anom48b}\\
&\langle|\psi^2(r)|\rangle_u=
{1\over 2}\left({u\over\xi
S}\right)^{1/2}\left[1+2\sqrt{uS\xi}k_d(r)\right]\ ,&\qquad r<l
\label{anom48c}
\end{eqnarray}
We see from
eqs.(\ref{anom48a}), (\ref{anom48b}), (\ref{anom48c})
that the eigenfunction normalization is dominated
by the region $r\sim\xi_{ef}$, where $\xi_{ef}\sim\sqrt{\xi/uS}\ll\xi$
plays the role of an effective localization length. In the region
$\xi_{ef}\ll r\ll\xi$ the wave intensity falls down as $1/r^2$, and
transits to the conventional localization behavior at $r\gg\xi$.
Therefore, the appearance of an anomalously high amplitude
$|\psi^2(0)|=u\gg 1/S\xi$ is  not just a local fluctuation, but
rather a kind of a cooperative phenomenon corresponding to existence
of a whole region $r\lesssim\xi_{ef}$ with an unusually large amplitude
$|\psi^2(r)|={1\over 2}\sqrt{u/\xi S}\sim 1/S\xi_{ef}$.

Let us emphasize once more that what we have calculated is the average
value $\langle|\psi^2(r)|\rangle$ of the eigenfunction intensity with
the condition $|\psi^2(0)|=u$. It is natural to ask now what are its
fluctuations. This question is addressed in section \ref{s3.2}. We will also
explain there what is the reason for the ``quasi-jump'' of
$\langle|\psi^2|\rangle$ from $|\psi^2(0)|=u$ to
$\langle|\psi^2(r)|\rangle={1\over 2}\sqrt{u/\xi S}$ at $r\sim l$.

\subsubsection{Metallic state ($L\ll\xi$).}
\label{s3.1.2}

We will assume that the observation point $r=0$ is located somewhere in the
bulk of the sample, so that both $\tau_-=L_-/\xi$ and $\tau_+=L_+/\xi$
are of the same order of magnitude as $L/\xi=\tau_-+\tau_+$:
$$
\tau_-,\tau_+\sim L/\xi=1/g\ll 1
$$
The distribution of eigenfunction intensities is given by
eqs.(\ref{anom11}), (\ref{anom14}), and its behavior in various ranges
of the variable $y=uV$ is indicated in
eqs.(\ref{adm15a})--(\ref{adm15c}).  We will study the structure of
the ALS responsible for the far asymptotics (\ref{adm15c}) in the
region $uV\gg g$. As is seen from eqs.(\ref{anom11}), (\ref{anom14}),
this corresponds to a large value of the argument
$t=2z^{1/2}=2(uS\xi)^{1/2}\gg g$ of the modified Bessel function in
eq.(\ref{anom14}). Under this condition, the integral in
eq.(\ref{anom14}) can be evaluated via the stationary point method
with use of the asymptotic expression(\ref{anom41}) for the modified
Bessel function:
\begin{eqnarray}
W^{(1)}(z=t^2/4,\tau)&=&{\tau\over\pi}\sqrt{{t\over
2\pi}}e^{-t}\int_0^\infty d\nu\exp\left\{{\pi\nu\over 2}-{\nu^2\over
4}\tau-tf\left({\nu\over t}\right)\right\} \nonumber\\
&=&{1\over\pi}\sqrt{2t\tau}\exp\left\{-t+{\pi^2\over 4\tau}
-t\tilde{f}\left({\pi\over t\tau}\right)\right\}\ ,\qquad
t\gg 1/\tau\gg 1;\label{anom49}\\
&& \tilde{f}(w)={w^2\over 2}+\ldots\ ,\qquad w\ll 1 \nonumber
\end{eqnarray}
The distribution function ${\cal P}(u)$ has therefore the form
\begin{eqnarray}
{\cal P}(u) &=
{16\over \pi^2}\sqrt{{S\xi\over u}}{\sqrt{L_+ L_-}\over
L} \exp & \left\{-4\sqrt{u\xi S}
+{\pi^2\xi\over 4L_+}
+{\pi^2\xi\over 4L_-}\right. \nonumber\\&& \left.
-2\sqrt{u\xi S}\left[
\tilde{f}\left({\pi\over 2 L_+}\sqrt{{\xi\over uS}}\right)
+\tilde{f}\left({\pi\over 2 L_-}\sqrt{{\xi\over uS}}\right)
\right]
\right\}
\nonumber\\
&={16\over \pi^2}\sqrt{{S\xi\over u}}{\sqrt{L_+ L_-}\over
L} \exp &\left\{-4\sqrt{u\xi S}
+{\pi^2\xi\over 4L_+}\left(1-{\sqrt{\xi/ uS}\over L_+}+\ldots\right)
\right.\nonumber\\&& \left.
+{\pi^2\xi\over 4L_-}\left(1-{\sqrt{\xi/ uS}\over L_-}+\ldots\right)
\right\}
\label{anom50}
\end{eqnarray}
This is just the formula (\ref{adm15c}), but written now with full
accuracy with respect to the subleading factors.

To calculate ${\cal Q}(u,r)$, eq.(\ref{anom35}), we have first
to evaluate the function $W^{(2)}(z,\tau_1,\tau_2)$,
eq.(\ref{anom33}). The contribution to it from the first term in
eq.(\ref{anom34}) was calculated in the preceding subsection, where
the insulating regime was considered. We will find that in the
metallic regime the second term in eq.(\ref{anom34}) gives a much
larger contribution, so that the first one can be
neglected. Substituting the second term of eq.(\ref{anom33}) into
(\ref{anom34}), we find after simple algebraic transformations:
\begin{eqnarray}
W^{(2)}(z=t^2/4,\tau_1,\tau_2)&=&{t\over 4\pi}\int d\nu K_{i\nu}(t)
e^{-{1+\nu^2\over 4}\tau_1} \nu\sinh(\pi\nu)
\nonumber \\ &\times&
\int{d\nu_1\,\nu_1\over
1+\nu_1^2} \sinh{\pi\nu_1\over
2}{\nu_1^2-\nu^2\over\cosh(\pi\nu_1)-\cosh(\pi\nu)}
e^{-{1+\nu_1^2\over 4}\tau_2}
\label{anom51}
\end{eqnarray}
Analysis of the double integral in eq.(\ref{anom51}) shows that
one should distinguish between the two possible siuations:

i) $\tau_1+{2\over t}\ll \tau_2^2\sim 1/g^2.$\\
This corresponds to a very large amplitude $u\sim{t^2\over\xi S}\gg
V^{-1}g^3$. The leading contribution to the integral in (\ref{anom51})
comes from the region $\nu\gg\nu_1$. To check this, let us assume that
$\nu\gg\nu_1$ and collect the  exponential factors in
eq.(\ref{anom51}):
$$
W^{(2)}(z=t^2/4,\tau_1,\tau_2)\sim\int
d\nu\exp\left\{-\nu^2{\tau_1\over 4}-{\nu^2\over 2t}\right\}\int
d\nu_1\exp \left\{{\pi\nu_1\over 2}-{\nu_1^2\tau_2\over 4}\right\}\ ,
$$
so that the characteristic values of the variables are
$\nu\sim(\tau_1+2/t)^{-1/2}$ and $\nu_1\sim 1/\tau_2$, confirming the
consistency of our assumption in the considered range of
parameters. Collecting all the prefactors, we get
\begin{equation}
W^{(2)}(z=t^2/4,\tau_1,\tau_2)={1\over 2\pi}\sqrt{{t\tau_+\over
2}}{1\over (1/t+\tau_1/2)^2}e^{\pi^2/4\tau_+-t}\
;\qquad\tau_+=\tau_1+\tau_2
\label{anom52}
\end{equation}
Note that we have omitted the corrections of the type
$t\tilde{f}(\pi/t\tau_+)$ in the exponent, since they are small in the
considered case $t\gg g^2$. Substituting eq.(\ref{anom52}) in
eq.(\ref{anom35}), we find
\begin{equation}
{\cal Q}(u,r)={8\over \pi^2 t^2}{\sqrt{L_+ L_-}\over L}
{1\over (1/t+r/2\xi)^2}\exp\left\{ {\pi^2\over 4}{\xi\over L_+}
+{\pi^2\over 4}{\xi\over L_-}-2t\right\}\ ;\qquad t=2\sqrt{uS\xi}\ ,
\label{anom53}
\end{equation}
and finally
\begin{equation}
\langle|\psi^2(r)|\rangle_u=
{1\over 2}\left({u\over\xi S}\right)^{1/2} {1\over
(1+r\sqrt{uS\over\xi})^2 }
\label{anom54}
\end{equation}

ii) $\tau_2^2\ll\tau_1+{2\over t}$.\\
In this case the integrals in (\ref{anom51}) are dominated by the
domain $\nu_1\sim\nu\gg 1$. Introducing the new variable
$\nu_-=\nu-\nu_1$, we reduce eq.(\ref{anom51}) to the form
\begin{eqnarray}
W^{(2)}(z=t^2/4,\tau_1,\tau_2)&=&{1\over 8}\sqrt{{t\over 2\pi}}\int d\nu
d\nu_- {\nu
\nu_-\over\sinh(\pi\nu_-/2)}
\nonumber \\
&\times&\exp\left\{-t-tf\left({\nu\over
t}\right)+{\pi\nu\over 2}-{\nu^2\over 4}\tau_1-{\nu_-^2\over
4}\tau_2+{\nu\nu_-\over 2}\tau_2\right\}
\nonumber
\end{eqnarray}
Evaluating the integrals, we find
\begin{eqnarray}
W^{(2)}(z=t^2/4,\tau_1,\tau_2)&=&{1\over 4\pi}\sqrt{{2t\over\tau_+^3}}
\left[\zeta\left(2,\:1-{\tau_1+2/t\over 2\tau_+}\right)+
\zeta\left(2,\:{\tau_1+2/t\over 2\tau_+}\right)\right]
\nonumber \\&&\hspace{4cm}\times\exp\left\{
{\pi^2\over 4\tau_+}-t\tilde{f}\left({\pi\over t\tau_+}\right)
\right\}\,
\label{anom55}
\end{eqnarray}
where $\zeta(p,z)$ is the generalized Riemann's zeta-function:
\begin{equation}
\zeta(p,z)=\sum_{k=0}^\infty(z+k)^{-p}
\label{anom56}
\end{equation}
Substituting this in eq.(\ref{anom35}), we get
\begin{eqnarray}
{\cal Q}(u)&=&{8\xi\over \pi^2 t^2L}\sqrt{{\tau_1\over\tau_+^3}}
\left[\zeta\left(2,\:1-{\tau_1+2/t\over 2\tau_+}\right)+
\zeta\left(2,\:{\tau_1+2/t\over 2\tau_+}\right)\right]
\nonumber \\ &&\hspace{2cm}\times
\exp\left\{
{\pi^2\over 4\tau_+}+{\pi^2\over 4\tau_-}
-2t-t\tilde{f}\left({\pi\over t\tau_+}\right)
-t\tilde{f}\left({\pi\over t\tau_-}\right)\right\}\ ,\qquad
t=2\sqrt{uS\xi}
\label{anom57}
\end{eqnarray}
Therefore, the average local intensity of the ALS is given by
\begin{equation}
\langle|\psi^2(r)|\rangle_u={1\over 8SL_+^2}\sqrt{{\xi\over uS}}
\left[\zeta\left(2,\:{r+\sqrt{\xi/ Su}\over 2L_+}\right)+
\zeta\left(2,\:1-{r+\sqrt{\xi/ Su}\over 2L_+}\right)
 \right]
\label{anom58}
\end{equation}
In particular, in its ``core'', $r\ll L_+$, the ALS intensity has
again the form (\ref{anom54}).

Comparing eqs.(\ref{anom54}),
(\ref{anom58}) with the result(\ref{anom48b}), we conclude that in its
central ``bump'' the ALS in the metallic sample has exactly the same
spatial structure as in a long (insulating) one.
The condition $Vu\gg g$, under which the asymptotical behavior
(\ref{adm15c}), (\ref{anom50}) is valid, acquires now a very
transparent meaning. This is just the condition that the effective
localization length of an ALS, $\xi_{ef}=\sqrt{\xi/uS}$ is much less
than the sample size $L$. Indeed,
$\xi_{ef}/L=\sqrt{\xi/uSL^2}=\sqrt{g/uV}$. In the ``tail'', $r\sim
L\gg \xi_{ef}$, the form of the ALS intensity is slightly modified by
the boundary of the sample, see eq.(\ref{anom58}).

All the above calculations in this subsection were valid for the
domain $r>l$. As to the region $r<l$, we can easily satisfy ourselves
using eqs.(\ref{anom37}) and (\ref{anom49}) that the result
(\ref{anom48c}) holds. Therefore, the ``quasi-jump'' of
$\langle|\psi^2(r)|\rangle_u$ at $r\ll l$  has the same form, as
in the insulating regime.

\subsection{Fluctuations of the eigenfunction of an anomalously
localized state.}
\label{s3.2}

In the section \ref{s3.1} we have calculated the average intensity
$\langle|\psi^2(r)|\rangle_u$ of a quantum state which has an
anomalously high amplitude $|\psi^2(0)|=u$ in a certain point
$\bbox{r}=0$. However, this average value does not give  full
information about the ALS. In the present subsection, we will study
the fluctuations of the intensity $|\psi^2(r)|$ of an ALS fixed by the
condition $|\psi^2(0)|=u$. Similarly to the subsection \ref{s3.1} (see
eq.(\ref{anom51})), we express the moments
$\langle |\psi(\bbox{r})|^{2p} |\psi(0)|^{2q}\rangle$
in terms of the Green's functions:
\begin{eqnarray}
&&\langle |\psi(\bbox{r})|^{2p} |\psi(0)|^{2q}\rangle \nonumber\\
&&\equiv {1\over \nu_{0} V}
\langle\sum_\alpha |\psi_\alpha(\bbox{r})|^{2p} |\psi_\alpha(0)|^{2q}
\delta(E-E_\alpha)\rangle \nonumber \\
&&=\lim_{\eta\to 0} {i^{q-p}\over 2\pi\nu_{0} V}
{(q-1)!(p-1)!\over(q+p-2)!}(2\eta)^{q+p-1} \langle
G_R^q(0,0) G_A^p(\bbox{r},\bbox{r})\rangle\ ,
\label{anom59}
\end{eqnarray}
The analogue of eq.(\ref{anom17}) reads
\begin{eqnarray}
G_R^q(0,0) G_A^p(\bbox{r},\bbox{r})&=& {i^{p-q}\over p!q!}\int D\Phi\,
D\Phi^\dagger (S_1(0) S_1^*(0))^q (S_2(\bbox{r})S_2^*(\bbox{r}))^p
\nonumber\\&\times &
\exp\left\{i\int d\bbox{r'} \Phi^\dagger(\bbox{r'}) \Lambda^{1/2}
(E+i\eta\Lambda-\hat{H})\Lambda^{1/2} \Phi(\bbox{r'})\right\}
\label{anom60}
\end{eqnarray}
Proceeding further as in the section \ref{s3.1}, we get for $r\gg l$
\begin{equation}
\langle |\psi(\bbox{r})|^{2p} |\psi(0)|^{2q}\rangle={(-1)^p\over
2V}\lim_{\eta\to 0} (2\pi\nu_{0}\eta)^{q+p-1}{(q-1)!(p-1)!\over(q+p-2)!}
\int DQ\,
Q_{11,bb}^q(0)Q^p_{22,bb}(\bbox{r})e^{-S\{Q\}}
\label{anom61}
\end{equation}
For the quasi--1D geometry  the integral can be again evaluated via the
method of \cite{FMRev}, yielding
\begin{equation}
\langle |\psi(r)|^{2p} |\psi(0)|^{2q}\rangle=
{1\over V(\xi S)^{q+p-1}} {q!p!\over(q+p-2)!}
\int_0^\infty dz\,z^{q-2} W^{(1)}(z,\tau_-)
W_p^{(2)}(z,\tau_1,\tau_2)\ ,
\label{anom62}
\end{equation}
where $W_p^{(2)}$ satisfies the same equation (\ref{anom31}) as the
function $W^{(2)}\equiv W_1^{(2)}$, and the boundary condition
generalizing eq.(\ref{anom32}):
\begin{equation}
W_p^{(2)}(z,0,\tau_2)=z^pW^{(1)}(z,\tau_2)
\label{anom63}
\end{equation}
Defining similarly to eq.(\ref{anom10}),
\begin{equation}
{\cal Q}_p(u,\bbox{r})= {1\over\nu_{0}
V}\langle\sum_\alpha |\psi_\alpha(\bbox{r})|^{2p}
\delta(|\psi_\alpha(0)|^2-u)\delta(E-E_\alpha)\rangle\ ,
\label{anom64}
\end{equation}
we get
\begin{equation}
{\cal Q}_p(u,r)={p\over V (S\xi)^p}\int _1^\infty {dv\over v}\left(1-{1\over
v}\right)^{p-1} {\partial^2\over \partial u^2}\left[W_p^{(2)}(vu\xi
S,\tau_1,\tau_2) W^{(1)}(vu\xi S,\tau_-)\right]\ ,\qquad p>1
\label{anom65}
\end{equation}
For the second moment, $p=2$, this formula can be simplified
\begin{equation}
{\cal Q}_2(u,r)=2\frac {W_2^{(2)}(u\xi
S,\tau_1,\tau_2) W^{(1)}(u\xi S,\tau_-)}
{V (uS\xi)^2}
\label{anom66}
\end{equation}
The solution to eqs.(\ref{anom31}), (\ref{anom63}) has the form
\begin{eqnarray}
&&W_p^{(2)}(z,\tau_1,\tau_2)=2\sqrt{z}\int_0^\infty d\nu\,
b_p(\nu,\tau_2)K_{i\nu}(2\sqrt{z}) e^{-{1+\nu^2\over 4}\tau_1}\ ;
\label{anom67}\\
&& b_p(\nu,\tau_2)={2\nu\sinh(\pi\nu)\over \pi^2}\int_0^\infty {dt\over
t^2} \left({t^2\over 4}\right)^p K_{i\nu}(t) W^{(1)}(t^2/4,\tau_2)
\label{anom68}
\end{eqnarray}

As we have seen above, an ALS in a metallic sample has essentially the
same spatial structure as in an insulating one. Physical reason for
this is very simple: an ALS with the effective localization length
$\xi_{ef}$ may exist in a sample with the length $L\gg\xi_{ef}$
without essential modifications. This reason is equally valid for the
fluctuations. Therefore, we can limit ourselves by studying the
fluctuations in the technically simpler case of an infinitely long
sample. In this case, the function $W^{(1)}$ is given by
eq.(\ref{anom38}), and eq.(\ref{anom68}) reduces to
\begin{eqnarray}
b_p(\nu,\tau_2=\infty)&=&{\nu\sinh(\pi\nu)\over 4\pi^2(2p-1)!}
\left|\Gamma\left(p+{1+i\nu\over 2}\right)\right|^2
\left|\Gamma\left(p-{1+i\nu\over 2}\right)\right|^2
\nonumber\\
&=&{\nu\tanh(\pi\nu/2)\over 2^{4p-1}(2p-1)!}
\left[\nu^2+(2p-1)^2\right]\left[\nu^2+(2p-3)^2\right]^2
\ldots \left[\nu^2+1\right]^2
\label{anom69}
\end{eqnarray}
The integral in eq.(\ref{anom67}) is determined by $\nu\gg
1$. Therefore, for not too  high moments $p$ the inequality
$p\ll \nu$ is satisfied; the corresponding restriction on $p$ will be
found explicitly below. We have then
\begin{equation}
b_p(\nu,\infty)\simeq{1\over (2p-1)!}\left({\nu\over
2}\right)^{4p-1}\ ;
\label{anom70}
\end{equation}
and consequently from eq.(\ref{anom67}),
\begin{equation}
W_p^{(2)}(z=t^2/4,\tau_1,\infty)\simeq e^{-t}\sqrt{{\pi t\over
2}}{1\over (\tau_1+2/t)^{2p}}
\label{anom71}
\end{equation}
Substituting this in eq.(\ref{anom65}), we get
\begin{equation}
{\cal Q}_p(u,r)\simeq {4\pi p\over
V}(S\xi)^{3/2-p}u^{-1/2}\int_1^{\infty}dv\, v^{1/2}\left(1-{1\over
v}\right)^{p-1} e^{-4\sqrt{vu\xi S}}{1\over\left(\tau_1+\sqrt{{1\over
vu\xi S}}\right)^{2p}}
\label{anom72}
\end{equation}
For not too large $p\ll\sqrt{u\xi S}$ the integral is determined by
the region $v-1\ll 1$, because of the factor
$\exp\{-4\sqrt{vu\xi S}\}$. It can be then estimated as
\begin{eqnarray}
{\cal Q}_p(u,r)&\simeq& {4\pi p\over V}(S\xi)^{3/2-p}u^{-1/2}
{1\over\left(\tau_1+\sqrt{{1\over u\xi S}}\right)^{2p}}
\int_1^\infty dv (v-1)^{p-1} e^{-4\sqrt{vu\xi S}}
\nonumber\\
&=& {4\pi\over V}p!(S\xi)^{3/2}u^{-1/2}e^{-4\sqrt{u\xi S}}
\left[{\sqrt{u/\xi S}\over 2(1+\tau_1\sqrt{uS\xi})^2}\right]^p
\label{anom73}
\end{eqnarray}
Finally, the $p$-th moment of the ALS intensity is given by
\begin{equation}
\langle|\psi^{2p}(r)|\rangle_u = \frac {{{\cal Q}_p}(u,r)}
{{\cal P}(u)} = p!
\left[{\sqrt{u/\xi S}\over 2(1+r\sqrt{uS/\xi})^2}\right]^p
\label{anom74}
\end{equation}

In course of the derivation we assumed that $p\ll\nu$ for typical
values of $\nu$ in the integral (\ref{anom67}). These values are
$$
\nu^2\sim{p\over\tau_1+{1\over\sqrt{uS\xi}}}\ ,
$$
so that the condition reads
\begin{equation}
p\ll {1\over {r\over\xi}+{1\over\sqrt{uS\xi}}}\equiv p_{max}
\label{anom75}
\end{equation}
An ALS is defined by the condition $uS\xi\gg 1$, so that for $r\ll\xi$
we have $p_{max}\gg 1$.

Comparing eq.(\ref{anom74}) with eq.(\ref{anom48b}), we find that in
this region
\begin{equation}
\langle|\psi^{2p}(r)|\rangle_u \simeq p!
\left[\langle|\psi^{2}(r)|\rangle_u\right]^p\ ,\ \ \ l<r\ll\xi,\ \
p\ll p_{max}
\label{anom76}
\end{equation}
This means that the fluctuations of the eigenfunction intensity
with respect to its average value are of the type
\begin{equation}
|\psi^2(r)|={\sqrt{u/\xi S}\over 2(1+r\sqrt{uS/\xi})^2}|\Phi^2(r)|\
,\qquad l<r\ll\xi
\label{anom77}
\end{equation}
where $|\Phi^2(r)|$ is distributed according to the GUE law
\begin{equation}
{\cal P}(|\Phi^2|)\simeq e^{-|\Phi^2|}
\label{anom78}
\end{equation}
The approximate result (\ref{anom78}) holds for not too large
$|\Phi^2|\ll p_{max}$.

Now we consider the fluctuations in the vicinity of the observation
point, $r\ll l$. Proceeding as in the section \ref{s3.1}, we find from
eqs.(\ref{anom59}), (\ref{anom60})
\begin{eqnarray}
\langle |\psi(r)|^{2p} |\psi(0)|^{2q}\rangle &=&
{1\over V (S\xi)^{p+q-1}}\sum_j {p!\over j!(p-j)!}{q!\over j!(q-j)!}
k_d^j(r){q!p!\over (q+p-2)!}\nonumber\\ &&\hspace{2cm}\times
\int dz\, z^{q+p-2}
W^{(1)}(z,\tau_+)W^{(1)}(z,\tau_-)
\label{anom79}
\end{eqnarray}
This allows one to restore the joint distribution function of
$|\psi(0)|^{2}$ and $|\psi(r)|^{2}$ (see Appendix), which looks
however rather cumbersome. For this reason, let us return to the
expression for the moments and consider $p=2$. We find then
\begin{equation}
\langle |\psi(r)|^{4} |\psi(0)|^{2q}\rangle =
{2\over  V(S\xi)^{q+1}}\left[1+2qk_d(r)+{q(q-1)\over 2}k_d^2(r)\right]
\int dz\,z^q
W^{(1)}(z,\tau_+)W^{(1)}(z,\tau_-)
\label{anom80}
\end{equation}
Restoring the function ${\cal Q}_2(u,r)$ defined in eq.(\ref{anom64}),
we get
\begin{equation}
{\cal Q}_2(u,r)={2\over V}\left[1-2k_d(r){d\over du}u+{k_d^2(r)\over 2}
{d^2\over du^2}u^2 \right] \{W^{(1)}(uS\xi,\tau_+)W^{(1)}(uS\xi,\tau_-)\}
\label{anom81}
\end{equation}
Substituting here (\ref{anom38}), we find
\begin{equation}
\langle|\psi^4(r)|\rangle_u\simeq k_d^2(r) u^2 +2u\sqrt{{u\over S\xi}}
k_d(r)(1-k_d(r))+ {u\over 2S\xi}(1-k_d(r))^2
\label{anom82}
\end{equation}
Therefore, the variance of $|\psi^2|$ is equal to
\begin{eqnarray}
\mbox{var}_u(|\psi^2|)&\equiv& \langle|\psi^4(r)|\rangle_u -
\langle|\psi^2(r)|\rangle_u^2 \nonumber\\
&\simeq&\left\{
\begin{array}{ll}
k_d(r)(1-k_d(r))u\sqrt{{u\over S\xi}}\ ,\qquad & r\ll r_0 \\
{u\over 4S\xi}\ , \qquad & r\gg r_0\ ,
\end{array}
\right.
\label{anom83}
\end{eqnarray}
with the scale $r_0$ as defined after eq.(\ref{anom47}). We find that
for $r\ll r_0$ the fluctuations are suppressed:
\begin{equation}
{\mbox{var}_u(|\psi^2(r)|) \over \langle|\psi^2(r)|\rangle_u^2}
\simeq {1-k_d(r)\over k_d(r)} {1\over\sqrt{uS\xi}}\ll 1
\label{anom84}
\end{equation}
On the other hand, for $r\gg r_0$ we have the GUE result
\begin{equation}
\mbox{var}_u(|\psi^2(r)|)\simeq \langle|\psi^2(r)|\rangle_u^2
\label{anom85}
\end{equation}
More generally, it is possible to check that for $r\gg r_0$ the terms
containing $k_d(r)$ become negligible in higher moments of
$|\psi(r)^2|$ as well, and the GUE-like fluctuations
(\ref{anom76})--(\ref{anom78}) take place.

Let us summarize the results of this subsection. We have found that
the ALS intensity $|\psi^2(r)|$ exhibits
in the region $r_0\ll r\ll\xi$ the GUE type fluctuations
(\ref{anom76})--(\ref{anom78}) with respect to its average value
$\langle|\psi^2(r)|\rangle_u$. These fluctuations are completely
analogous to those for an ordinary delocalized state in a metallic
sample, see eq.(\ref{anom2u}). The difference is that in the present
case the average intensity $\langle|\psi^2(r)|\rangle_u$ is not uniform
in the coordinate space.
In the region  $r\ll r_0$ the fluctuations are suppressed:
$\mbox{var}_u(|\psi^2(r)|)\ll \langle|\psi^2(r)|\rangle_u^2$. This is
similar to what has been found in \cite{prigodin} for the spatial
structure of an ``ordinary'' delocalized state with a moderately large
local intensity $u=|\psi^2(0)|$ (when the zero-dimensional formula
(\ref{anom2u}) holds and an ALS is not formed).
This means that the intensity of a typical ALS is in this region close
to the average value $\langle|\psi^2(r)|\rangle_u$, which exhibits the sharp
decrease from $|\psi^2(0)|=u$ to
$\langle|\psi^2(r)|\rangle_u={1\over 2}\sqrt{{
u\over\xi S}}$ at $r\gg r_0$. It is not difficult to understand that
 this quasi-jump  has the same sourse as the GUE-like fluctuations at
$r\gg r_0$. One can ask, of course, why this  short-scale
fluctuation happens exactly in the center of the smooth ALS ``bump''
with a probability close to unity. The answer is as follows. We are
studying the states with an anomalously large local intensity $u$,
which is an exponentially rare event. There are two sources which
may favor the formation of such a high intensity: i) formation of an
ALS with a spatially non-uniform smooth envelope, and ii) short-scale
GUE-like fluctuations. Both these mechanisms have exponentially small
probabilities to produce an enhancement of the intensity by a large
factor. The found configuration of $\langle|\psi^2(r)|\rangle_u$ (short-scale
quasi-jump (\ref{anom48c}) on top of the smooth configuration
(\ref{anom48b})) represents just the optimal combination of the two
mechanisms.

\section{Generalization to higher spatial dimensions.}
\label{s4}
\setcounter{equation}{0}
The calculation in the preceding section relies essentially on the
quasi-1D structure of the sample, so it is not applicable for
a sample of higher dimensionality.
In this case, the asymptotical behavior of the distribution
function ${\cal P}(u)$ was studied by Fal'ko and Efetov
\cite{falef}. These authors used eqs.(\ref{4}), (\ref{4o}) and applied
the saddle-point method suggested for the supersymmetric
$\sigma$--model by Muzykantskii and Khmelnitskii \cite{mk}. The
advantage of this method is that it is applicable for any spatial
dimension $d$. It was conjectured in \cite{mk,falef} that the saddle
point solution mimics the spatial form of the ALS. Having at our
disposal the exact solution for the quasi-1D case, we can check the
accuracy of the saddle-point method and of this conjecture. Comparing
the form of ${\cal P}(u)$ found in the quasi-1D case by the saddle-point
method \cite{falef} with the result of the exact solution (\ref{anom50}),
we find a very good agreement between them. Furthermore, the
saddle-point configuration in the quasi-1D case has the form
\cite{falef,ldos-as}
\begin{equation}
{\lambda_1(r)\over\lambda_1(0)}\equiv e^{\theta(r)-\theta(0)}=
{1\over \left(1+r\sqrt{{u\over 2\pi\nu_{0} D}}\right)^2}\ ;\qquad 0<r\ll
L_+
\label{anom86}
\end{equation}
Here $\lambda_1=e^{\theta}/2$ is the eigenvalue of the $Q$-matrix which
has been introduced after eq.(\ref{1}). Comparing eq.(\ref{anom86})
with eqs.(\ref{anom48b}), (\ref{anom54}), we see that the saddle-point
solution nicely reproduces the average intensity of the ALS,
$\langle|\psi^2(r)|\rangle_u$ for $r>l$, up to an overall normalization
factor. Being encouraged by this agreement, we will now use the
results of the saddle-point study to describe the structure of the ALS
in $d=2,3$.

\subsection{2D geometry.}
\label{s4.1}

For a 2D disk-shaped sample of a radius $L$
with the high amplitude point
$\bbox{r}=0$ in the center of the disk, the saddle-point solution was
found to have the form \cite{falef}
\begin{equation}
e^{\theta(r)-\theta(0)}=\left({r\over l_*}\right)^{-2\mu}
\left\{1-{l_*^2 u\over 8(1-\mu)^2\pi\nu_{0} D}
\left({r\over l_*}\right)^{2-2\mu}\right\}^{-2}
\approx \left({r\over l_*}\right)^{-2\mu}\ ,\qquad r\ge l_*
\label{anom87}
\end{equation}
where the exponent $0<\mu<1$ depends on $u$ and  satisfies the equation
\begin{equation}
\left({L\over l_*}\right)^{2\mu}={2-\mu\over 8\mu(1-\mu)^2} {L^2
u\over\pi\nu_{0} D}
\label{anom88}
\end{equation}
We are interested in the asymptotic region $uL^2\gg \pi\nu_{0}
D\ln^{-1}(L/l)$, where the distribution of the eigenfunction intensity
is given by eq.(\ref{anom7}), and an ALS is formed. Then the exponent
$\mu$ can be approximated as
\begin{equation}
\mu\simeq{\ln\left({L^2 u\over 2\pi\nu_{0} D}\ln{L\over
l_*}\right)\over 2\ln(L/ l_*) }
\label{anom88a}
\end{equation}
The lower cut-off scale $l_*$ appears in eq.(\ref{anom87}) because of the
restriction of the diffusion approximation on the momenta $q$ of the
$\sigma$-model field: $q<l^{-1}$. Correspondingly, it is determined by
the condition \cite{mk,relax,ldos-as}
$$
{d\over dr}\theta(r)|_{r=l_*}\sim l^{-1}\ ,
$$
which yields $l_*\sim\mu l$, so that $l_*$ differs from $l$ by the
logarithmic factors (\ref{anom88a}) only.

Normalizing properly the expression (\ref{anom87}), we find that the
average ALS density for $r>l_*$ is equal to
\begin{equation}
\langle|\psi^2(r)|\rangle_u={u\over 4\pi^2\nu_{0} D\mu}\left({r\over
l_*}\right)^{-2\mu}
\left\{1-{l_*^2 u\over 8(1-\mu)^2\pi\nu_{0} D}
\left({r\over l_*}\right)^{2-2\mu}\right\}^{-2}
\ ,\qquad r\ge l_*
\label{anom89}
\end{equation}
The saddle-point calculation of Ref.\cite{falef}
assumes that $\theta(r)$ is constant for
$r<l_*$, so that eq.(\ref{anom89}) gives
$\langle|\psi^2(r)|\rangle_u\simeq{u\over 4\pi^2\nu_{0}D\mu}$ in this
region.  However, for very small $r<l$ the average intensity
$\langle|\psi^2(r)|\rangle_u$ changes
sharply, as we have seen in the quasi--1D case. Indeed, in this domain
the function ${\cal Q}(u,r)$ is given by eq.(\ref{anom36}), where
$Y_a(u)$ has the following behavior \cite{falef}
\begin{equation}
Y_a(u)\sim\exp\left\{-\pi^2\nu_{0} D
{\ln^2\left({V u\over 2\pi^2\nu_{0} D}\ln{L\over
l_*}\right)\over\ln(L/l_*) } \right\}
\label{anom90}
\end{equation}
Using now eq.(\ref{anom36}) for ${\cal Q}(u,r)$ and eq.(\ref{4}) for
${\cal P}(u)$ , we get
\begin{equation}
\langle|\psi^2(r)|\rangle_u\equiv {{\cal Q}(u,r)\over {\cal P}(u)}=
[1-k_2(r)+A(u)k_2(r)]\langle|\psi^2(r=l_*)|\rangle_u\ ,\qquad r<l_*
\label{anom91}
\end{equation}
Here the height of the quasi-jump is given by
\begin{equation}
A(u)\simeq -u{d\over du}\ln Y_a(u)\simeq {2\pi^2\nu_{0} D\over
\ln(L/l_*)}
\ln\left({Vu\over 2\pi^2\nu_{0} D}\ln{L\over l_*}\right)\simeq 4\pi^2\nu_{0}
D\mu
\label{anom92}
\end{equation}
Combining eqs.(\ref{anom89}), (\ref{anom91}), and  (\ref{anom92}), we
get
\begin{equation}
\langle|\psi^2(r)|\rangle_u={u\over A(u)}[1-k_2(r)+k_2(r)A(u)]\ ,\qquad
r<l_*
\label{anom93}
\end{equation}
In particular, at $r=0$ we find $\langle|\psi^2(r)|\rangle_u=u$, that
shows the perfect consistency of the whole procedure.

\subsection{3D geometry.}
\label{s4.2}
Now, we consider the 3D case. The saddle-point configuration has the
form \cite{falef,ldos-as}
\begin{equation}
\theta(r)-\theta(0)=C\left({l_*\over r}-1\right)\ ,\qquad r\ge l_*\ ,
\label{anom94a}
\end{equation}
where the coefficient $C$ is defined by the condition
\begin{equation}
Ce^C\sim{uV\over\nu_{0}Dl_*}\ ,
\label{anom94b}
\end{equation}
and $l_*=Cl$ is the cut-off length scale which has the same origin as
in 2D. We are interested in the asymptotic region $uV\gg\nu_{0}Dl$,
where eq.(\ref{anom94b}) yields
\begin{eqnarray}
e^C&\sim&{uV\over g(l)\ln^2[uV/g(l)]}\ , \label{anom94c}\\
l_*&\sim&l\ln[uV/g(l)]\ , \label{anom94d}
\end{eqnarray}
where  $g(l)\sim(k_Fl)^2$ is the conductance on the scale of order of
the mean free path $l$. The optimal configuration(\ref{anom94a}) has
the form
\begin{equation}
e^{\theta(r)-\theta(0)}=\left\{{uV\over g(l)\ln^2[uV/g(l)]}
\right\}^{{l_*/ r}-1}\
,\qquad r\ge l_*
\label{anom94e}
\end{equation}
This configuration
determines the asymptotical behavior of the functions $Y_a(u)$, $\cal
P(u)$ at $Vu\gg g(l)$:
\begin{equation}
{\cal P}(u)\sim Y_a(u)\sim\exp\left\{-\mbox{const}\: g(l)\ln^3{Vu\over
g(l)}\right\}
\label{anom95}
\end{equation}
Note that in order to fix the numerical coefficient in
eq.(\ref{anom95}), one has to go beyond the long-wave-length
$\sigma$-model approximation. Such a ``ballistic'' generalization of
the $\sigma$-model was recently suggested in Ref.\cite{mk2}.
The height of the quasi-jump $A(u)$ is  found from
eq.(\ref{anom92}) to be
\begin{equation}
A(u)\sim g(l)\ln^2{Vu\over g(l)}
\label{anom96}
\end{equation}
Thus, the density of ALS is
\begin{equation}
\langle|\psi^2(r)|\rangle_u={u e^{\theta(r)-\theta(0)}\over A(u)}
\sim {1\over V}\left[{uV\over g(l)\ln^2 {uV\over g(l)}}\right]^{l_*/r}\
,\qquad r\ge l_*
\label{anom97}
\end{equation}
The quasi-jump at small $r$ is given again by eq.(\ref{anom93}), with
$k_2(r)$ replaced by $k_3(r)$.

The value of $\langle|\psi^2(r)|\rangle_u$ given by eq.(\ref{anom97})
exceeds considerably the average density of a delocalized state,
$\langle|\psi^2|\rangle=1/V$ at $r\lesssim l_{**}$, where
$$
l_{**} \sim l_*\ln{uV\over g(l)}\sim l \ln^2{uV\over g(l)}.
$$
For larger distances, $r\gtrsim l_{**}$, we have
$\langle|\psi^2(r)|\rangle_u\simeq 1/V$. Therefore, in contrast to
quasi-1D and 2D systems, where an ALS formation is a redistribution of
the eigenfunction intensity in the whole sample, in 3D it constitutes
just a ``bump'' with the extent of order of $l_{**}$, on top of the
usual background density $\langle|\psi^2|\rangle=1/V$.

\subsection{Fluctuations.}
\label{s4.3}
Let us now discuss briefly the fluctuations of the ALS intensity.
At small $r\ll l$, we find in full analogy with
eqs.(\ref{anom82})--(\ref{anom85}):
\begin{equation}
\langle|\psi^4(r)|\rangle_u\simeq u^2\left[k_d^2(r)+{4\over
A(u)}(1-k_d(r)) k_d(r)+{2\over A^2(u)}(1-k_d(r))^2\right]\ ,
\label{anom98}
\end{equation}
and consequently,
\begin{equation}
{\mbox{var}_u(|\psi^2(r)|) \over \langle|\psi^2(r)|\rangle_u^2}
\simeq \left\{
\begin{array}{ll}
2\displaystyle{ {1-k_d(r)\over k_d(r)A(u)} }\ ,\qquad & r\ll r_0\\
1\ ,\qquad & r\gg r_0
\end{array}
\right.
\label{anom99}
\end{equation}
Here $r_0$ is the characteristic spatial scale of the quasi-jump,
determined by the condition $k_d(r)A(u)\sim 1$. In 2D, using
eqs.(\ref{anom28}) and (\ref{anom92}), we find
\begin{equation}
r_0\sim l {\ln\left({Vu\over 2\pi^2\nu_{0} D}\ln (L/l)\right)\over
\ln(L/l)}
\label{anom100}
\end{equation}
In 3D the analogous calculation would give $r_0\sim l
\ln[Vu/g(l)]>l$. This means that in fact $r_0\sim l$, because
of the exponential decrease of $k_d(r)\sim e^{-r/l}$ at $r>l$, which
was not taken into account in eqs.(\ref{anom28}), (\ref{anom29}).
We see from eq.(\ref{anom99}) that at $r\ll r_0$ the fluctuations are
relatively weak,
$\mbox{var}_u(|\psi^2(r)|) \ll \langle|\psi^2(r)|\rangle_u^2$. In the
opposite limit, $r\gg r_0$, the fluctuations are expected to have
essentially the GUE statistics, similarly to what we have found for the
quasi-1D sample geometry, see eqs.(\ref{anom76})--(\ref{anom78}).

\section{Asymptotic behavior of distributions of various quantities,
and their interrelationship.}
\label{s5}
\setcounter{equation}{0}

In the preceding sections, we have considered the spatial structure of
ALS responsible for the asymptotic behavior of the distribution of
eigenfunction intensity, ${\cal P}(u)$. One can consider, however, other
quantities, large values of which indicate in some sense a stronger
localization of an eigenfunction, so that the asymptotic behavior of
corresponding distribution function is also related to a kind of ALS.
An illustrative
example of such a quantity is the inverse partcipation ratio (IPR)
$I_2=\int d^dr|\psi^4(\bbox{r})|$. Let us consider the case of an
infinite quasi-one-dimensional sample, for which the distribution
function ${\cal P}(I_2)$ was calculated exactly in Ref.\cite{mf7,FMRev}.
It has the following asymptotic behavior:
\begin{equation}
{\cal P}(I_2)\sim\exp\left\{-{\pi^2\over 2}\xi S I_2\right\}\ ,
\label{anom101}
\end{equation}
where we omitted  preexponential factors. Let us suppose that the
asymptotic form of ${\cal P}(I_2)$ is determined by the same anomalously
localized sates, which control the asymptotic behavior of $\cal
P(u)$. The intensity of such an ALS was found in Sec.\ref{s3} to be
\begin{equation}
|\psi^2(\bbox{r})|=|\psi^2(r)|_{\mbox{smooth}} |\Phi^2(\bbox{r})|\
,\qquad r\ll\xi
\label{anom102}
\end{equation}
where
\begin{equation}
|\psi^2(r)|_{\mbox{smooth}}={1\over 2S} {\xi_{ef}\over
(r+\xi_{ef})^2}\ ,\qquad \xi_{ef}=\sqrt{\xi/uS}\ ,
\label{anom103}
\end{equation}
and $|\Phi^2(\bbox{r})|$ exhibits the GUE-like fluctuations
(\ref{anom78}). The distribution ${\cal P}(u)$ behaves asymptotically as
${\cal P}(u)\sim\exp\{-4\sqrt{uS\xi}\}$; here the factor
$\exp\{-2\sqrt{uS\xi}\}$ is the GUE probability of the quasi-jump in the
vicinity of $r=0$ with $|\Phi^2(0)|=2\sqrt{uS\xi}$. The remaining
$\exp\{-2\sqrt{uS\xi}\}$ factor is therefore the weight of the envelope
configuration $|\psi^2(r)|_{\mbox{smooth}}$. The corresponding IPR
is equal to $I_2=(1/3)\sqrt{u/S\xi}$. Thus, assuming that this ALS
determines the asymptotics of the IPR distribution ${\cal P}(I_2)$, we
would get
\begin{equation}
{\cal P}(I_2)\sim\exp\left\{-6\xi S I_2\right\}.
\label{anom104}
\end{equation}
This result has the same exponential form, as the correct asymptotics
(\ref{anom101}), but with a larger numerical coefficient in the
exponent.  The explanation
for this discrepancy is the following: an ALS, which is optimal for
maximizing the local amplitude $u=|\psi^2(0)|$, does not optimize the
IPR. A detailed study, which will be published elsewhere \cite{mf8},
shows that the ALS, which determine the asymptotics of the IPR
distribution, have the following spatial shape:
\begin{equation}
 |\psi^2(r)|_{\mbox{smooth}}={1\over \pi S} {\xi_{ef}\over
r^2+\xi_{ef}^2}\ ,\qquad \xi_{ef}={1\over \pi I_2 S}\ ,
\label{anom105}
\end{equation}
that is different from eq.(\ref{anom103}). Thus, the spatial form of
an ALS depends on the specific physical quantity (local amplitude,
IPR, \ldots), for which it represents an optimal fluctuation. In the
remaining part of this section, we consider from this point of view
the distributions of LDOS and of the relaxation times in open metallic
samples, which have been studied via the supersymmetry approach in
Refs.\cite{mk,relax} and Ref.\cite{ldos-as}, respectively.

\subsection{Quasi-1D geometry.}
\label{s5.1}
\subsubsection{Distribution of relaxation times.}
\label{s5.1.1}
The long-time dispersion of
the average conductance has the LN form \cite{mk}
\begin{equation}
G(t)\sim\exp\left\{-g\ln^2{t\Delta\over \ln (t\Delta)}\right\}\ ;
\qquad t\gg \Delta^{-1}\ ,
\label{anom106}
\end{equation}
where $\Delta=1/\nu_{0} LS$ is the mean level spacing
and $g=2\pi\nu_{0} DS/L$
is the dimensionless conductance. We represent eq.(\ref{anom106}) as a
superposition of the simple relaxation processes with mesoscopically
distributed relaxation times \cite{akl}:
\begin{equation}
G(t)\sim\int dt_\phi e^{-t/t_\phi} {\cal P}(t_\phi)
\label{anom106a}
\end{equation}
The distribution function ${\cal P}(t_\phi)$ then behaves as follows:
\begin{equation}
{\cal P}(t_\phi)\sim\exp\{-g\ln^2(g\Delta t_\phi)\}\ ;\qquad t_\phi\gg
{1\over g\Delta}
\label{anom107}
\end{equation}
This can be easily checked by substituting eq.(\ref{anom107}) into
eq.(\ref{anom106a}) and calculating the integral via the sationary
point method; the stationary point equation being
\begin{equation}
2gt_\phi\ln(g\Delta t_\phi)=t
\label{anom108}
\end{equation}

Note that $t_D^{-1}\equiv g\Delta$ is the inverse characteristic time of
diffusion through the sample (Thouless energy), or in other
words, the typical width of a level of an open system. The formula
(\ref{anom107}) concerns therefore the states with anomalously small
widths in the energy space. The corresponding saddle-point
configuration is found from the requirment of providing a minimum to
the action
\begin{equation}
\cal S=-{\pi\nu_{0} D\over 4}\int d^dr\mbox{Str}(\nabla Q)^2={\pi\nu_{0}
D\over 2}\int d^dr(\nabla\theta)^2\ ,
\label{anom108a}
\end{equation}
with the additional restriction
\begin{equation}
\int d^dr(\cosh\theta-1)={t\over\pi\nu_{0}}\ ,
\label{anom108b}
\end{equation}
and the boundary condition
on the boundary with leads $\theta|_{\mbox{leads}}=0$. For a
quasi-1D sample of the length $L$, the solution can be approximated as
\cite{mk}:
\begin{equation}
\theta(r)\simeq\theta_0\left(1-{2|r|\over L}\right)\ ,\qquad -{L\over
2}<r<{L\over 2}\ ,
\label{anom109}
\end{equation}
where $\theta_0$ satisfies the equation
\begin{equation}
e^{\theta_0}={2\over\pi}t\Delta\theta_0\ ,
\label{anom110}
\end{equation}
so that
\begin{equation}
\theta_0\simeq\ln\left({2\over\pi}t\Delta\ln (t\Delta)\right)\simeq
\ln\left({4\over\pi}g\Delta t_\phi \ln^2(g\Delta t_\phi)\right).
\label{anom111}
\end{equation}
 Relying on our
previous experience, we believe that the smoothed intensity of the
corresponding state is $|\psi^2(r)|_{\mbox{smooth}}\propto
e^{\theta(r)}$, as was stated in Ref.\cite{mk}. Normalizing it by the
condition $\int |\psi^2(\mbox{r})|=1$, we get
\begin{equation}
|\psi^2(r)|_{\mbox{smooth}}={\theta_0\over SL}e^{-2\theta_0|r|/L}
\label{anom112}
\end{equation}
Thus, the ALS, which gives a minimum to the level width $t_\phi^{-1}$,
has an exponential shape (\ref{anom112}), (\ref{anom111}).

\subsubsection{Distribution of local density of states.}
\label{s5.1.2}
Now, let us consider the distribution ${\cal P}(\rho)$ of
LDOS. Typically, in an open metallic sample the LDOS
$\rho(E,\bbox{r})$ is given by a superposition of $\sim 1/t_D\Delta=g$
adjacent levels, since their widths are of order of $1/t_D$. However,
we can expect that for $\rho$ much greater than its average value
$\nu_{0}$, the asymptotic form of ${\cal P}(\rho)$ is determined by a
probability to have a single narrow resonance, which gives this value
 of LDOS $\rho(E,\bbox{r})$. The most favorable situation happens when the
resonance is located around the point $\bbox{r}$ in the real space and
around the energy $E$ in energy space. The LDOS provided by such a
resonance is expected to be:
\begin{equation}
\rho_{ALS}=|\psi^2(\bbox{r})|{2 t_\phi\over\pi}\ ,
\label{anom113}
\end{equation}
where $t_\phi^{-1}$ is the resonance width.
Thus, the optimal fluctuation shoul provide now a maximum to the
product of the local amplitude $u=|\psi^2(\bbox{r})|$ and the inverse
level width $t_\phi$. Since the distribution ${\cal P}(t_\phi)$,
eq.(\ref{anom107}), decays much slower than ${\cal P}(u)$,
eq.(\ref{anom42}), one should expect the asymptotic behavior of $\cal
P(\rho)$ to be mainly determined by ${\cal P}(t_\phi)$. Indeed, it was
found in Ref.\cite{ldos-as} that ${\cal P}(\rho)$ has a LN form, similar
to that of ${\cal P}(t_\phi)$:
\begin{equation}
{\cal P}(\rho)\sim\exp\left\{-{\xi\over 4}\left({1\over L_+}+{1\over
L_-}\right)\ln^2(\rho/\nu_{0})\right\}\ ,
\label{anom114}
\end{equation}
where, as before, $L_+$ and $L_-$ are the distances from the
observation point $r=0$ to the sample edges. The corresponding
saddle-point configuration reads:
\begin{equation}
e^{\theta(r)}\simeq\left\{
\begin{array}{ll}
(\rho/\nu_{0})^{1-r/L_+}\ ,&\qquad r>0\\
(\rho/\nu_{0})^{1-|r|/L_-}\ ,&\qquad r<0
\end{array}
\right.
\label{anom115}
\end{equation}
If we put the observation point in the middle of the sample,
$L_+=L_-=L/2$, the configuration (\ref{anom115}) acquires the same
form as the optimal configuration (\ref{anom109}) for the relaxation
time $t_\phi$. The corresponding values of $t_\phi$ and $\rho$ are
related as follows:
\begin{equation}
{4\over \pi}g\Delta t_\phi\ln^2(g\Delta t_\phi)=\rho/\nu_{0}\ ,
\label{anom116}
\end{equation}
or, expressing $t_\phi$ through $\rho$,
\begin{equation}
t_\phi={\pi\rho\over 4g\Delta\nu_{0}\ln^2(\rho/\nu_{0})}
\label{anom117}
\end{equation}
Now, we calculate the value of the local amplitude $|\psi^2(0)|$
for an ALS corresponding to the cofiguration (\ref{anom115}). First,
its smoothed intensity is given by
\begin{equation}
|\psi^2(r)|_{\mbox{smooth}}={\cal N}^{-1}
e^{\theta(r)}={\ln(\rho/\nu_{0})\over V}
\left({\rho\over\nu_{0}}\right)^{-2|r|/L}.
\label{anom118}
\end{equation}
Second, the quasi-jump induced by the GUE-type fluctuations gives an
additional factor, which can be found in the same way as prescribed by
eq.(\ref{anom36}):
\begin{equation}
A(\rho)=-\rho{\partial\over\partial\rho}\ln \cal
P(\rho)=2g\ln(\rho/\nu_{0})
\label{anom119}
\end{equation}
Combining eqs.(\ref{anom117}), (\ref{anom118}), and (\ref{anom119}),
we can compute the LDOS (\ref{anom113})
determined by this resonance state:
\begin{equation}
\rho_{ALS}(E,0)=|\psi^2(0)|_{\mbox{smooth}}\cdot A(\rho) \cdot
{2t_\phi\over\pi}=
{\ln(\rho/\nu_{0})\over V}\cdot 2g\ln(\rho/\nu_{0})\cdot {\rho V\over
2g\ln^2(\rho/\nu_{0})}=\rho
\label{anom120}
\end{equation}
We have explicitly checked therefore that the LDOS $\rho$ is indeed
determined by a single ALS, smoothed intensity of which is given by
eq.(\ref{anom118}). There are three sources of the enhancement of
LDOS: i) amplitude of the smooth envelope of the wave function, ii) the
short-scale GUE ``bump'', and iii) the inverse resonance width. They
are represented by the three factors in eq.(\ref{anom120}),
respectively. Note that the calculated $\rho_{ALS}$ reproduces the
value of $\rho$ with an amazing accuracy (including logarithmic
factors and even the numerical coefficient).

\subsubsection{Distribution of global density of states.}
\label{s5.1.3}

Finally, we discuss the contribution of ALS to the asymptotical
behavior of the distribution function ${\cal P}({\nu})$ of the
global density of states (DOS),
\begin{equation}
{\nu}(E)={1\over
V}\langle\sum_\alpha\delta(E-E_\alpha)\rangle={1\over V}\int
d^dr\rho(E,r)
\label{anom121}
\end{equation}
A resonance state with an energy $E$ and width $t_\phi^{-1}$ gives a
following contribution to ${\nu}(E)$:
\begin{equation}
{\nu}_{ALS}(E)={2\over \pi}{t_\phi\over
V}={2\over\pi}t_\phi\Delta\nu_{0}
\label{anom122}
\end{equation}
Thus, if we assume that the asymptotic behavior of ${\cal P}({\nu})$ is
determined by isolated (in energy space) anomalously localized states,
it will have the form:
\begin{equation}
{\cal P}({\nu})\sim {\cal P}\left(t_\phi={\pi{\nu}\over
2\Delta\nu_{0}}\right)\sim \exp\left\{-g\ln^2(g{\nu}/\nu_{0})\right\}.
\label{anom123}
\end{equation}
We will see below that an analogous procedure in 2D leads to a result
for ${\cal P}({\nu})$ which is in full agreement with the
renormalization group calculation of Altshuler, Kravtsov and Lerner
\cite{akl}.

\subsection{2D geometry.}
\label{s5.2}
\subsubsection{Distribution of relaxation times.}
\label{s5.2.1}

Distribution of relaxation times $t_\phi$ has the form \cite{relax}
\begin{equation}
{\cal P}(t_\phi)\sim\left\{
\begin{array}{ll}
(t_\phi/t_D)^{-4\pi g}\ , &\ \
t_D\ll t_\phi\ll t_D \left({L\over l}\right)^2 \\
\exp\left\{-{\pi g\over 2} {\ln^2(t_\phi/\tau)\over \ln
(L/l)}\right\} \ ,&\ \ t_\phi\gg t_D \left({L\over l}\right)^2\ ,
\end{array}
\right.
\label{anom124}
\end{equation}
with $g=2\pi\nu_{0}D$ being the dimensionless conductance of a 2D square.
An ALS corresponding to the first regime,
$t_D\ll t_\phi\ll t_D ({L/ l})^2$, has the following spatial
structure:
\begin{equation}
|\psi^2(r)|_{\mbox{smooth}}={\cal N}^{-1}e^{\theta(r)}=
{1\over 16\pi D t_\phi}{1\over [(r/L)^2+L^2/(16Dt_\phi)]^2}\ ,
\label{anom125}
\end{equation}
so that it has an effective localization length
$\xi_{ef}\sim L(t_D/t_\phi)^{1/2}$, with the intensity decreasing
 as $1/r^4$ outside
the region of the extent $\xi_{ef}$. As to the ultra-long-time region,
$t_\phi\gg t_D ({L/ l})^2$, the saddle-point solution reads:
\begin{equation}
e^{\theta(r)}={(r/L)^{\gamma_t-2}\over \left[(r/L)^{\gamma_t}+
{\gamma_t+2\over\gamma_t-2}(l_*^{(t)}/L)^{\gamma_t}\right]^2}\ ;\qquad
l_*^{(t)}\le r\le L\ ,
\label{anom125a}
\end{equation}
where
\begin{equation}
l_*^{(t)}=\gamma_t l\ ;\qquad
\gamma_t\simeq{\ln(t_\phi/t_D)\over\ln(R/l)}
\label{anom125b}
\end{equation}
Now, $l_t$ plays a role of an effective localization length, and the
intensity shows the following behavior:
\begin{equation}
|\psi^2(r)|\sim{1\over l_*^{(t)2}}\left({r\over
l_*^{(t)}}\right)^{-\gamma_t-2} \ , \qquad l_*^{(t)}\le r\le L.
\label{anom125c}
\end{equation}

\subsubsection{Distribution of local density of states.}
\label{s5.2.2}
Distribution of LDOS, ${\cal P}(\rho)$, has the following asymptotics
\cite{ldos-as}:
\begin{equation}
{\cal P}(\rho)\sim\exp\left\{-\frac{\pi^2\nu_{0} D \ln^2\rho}
{\ln (L/l^{(\rho)}_*)}\right\}.
\label{anom126}
\end{equation}
 The corresponding saddle-point solution reads
\begin{equation}
e^{\theta(r)}\simeq{\rho\over\nu_{0}}\left({l^{(\rho)}_*\over
r}\right)^{\gamma_\rho}\ ,
\label{anom127}
\end{equation}
where $l^{(\rho)}_*=\gamma_\rho l$, and
$$
\gamma_\rho={\ln(\rho/\nu_{0})\over \ln (L/l^{(\rho)}_*) }.
$$
Normalizing it, we get the following ALS intensity at $r\ge l^{(\rho)}_*$:
\begin{equation}
|\psi^2(r)|_{\mbox{smooth}}\simeq\left\{
\begin{array}{ll}
\displaystyle{
(2-\gamma_\rho){1\over V}{\rho\over\nu_{0}}\left({l^{(\rho)}_*\over
r}\right)^{\gamma_\rho}\ ;  }
  &\qquad \gamma_\rho<2\\
\displaystyle{
(\gamma_\rho-2){1\over \pi l_*^{(\rho)2}}\left({l^{(\rho)}_*\over
r}\right)^{\gamma_\rho}\ ;  }&\qquad \gamma_\rho>2
\end{array}
\right.
\label{anom129}
\end{equation}
The value of the GUE-type quasi-jump is
\begin{equation}
A(\rho)=-\rho{\partial\over\partial\rho}\ln\cal
P(\rho)=2\gamma_\rho\pi^2\nu_{0} D
\label{anom130}
\end{equation}
To estimate the escape time from this resonance state, we note that
its power-law decay, $|\psi^2(r)|\propto(r/l^{(\rho)}_*)^{-\gamma_\rho}$, is
similar to that of the ALS optimizing the relaxation time,
$|\psi^2(r)|\propto(r/l^{(t)}_*)^{-\gamma_t-2}$.
This allows us to identify
$\gamma_\rho=\gamma_t+2$, so that
\begin{equation}
t_\phi/t_D\sim\left\{
\begin{array}{ll}
\displaystyle{
{\rho\over\nu_{0}}\left({l\over L}\right)^2\ },&\qquad \gamma_\rho>2\\
1\ ,   &\qquad \gamma_\rho<2\ ,
\end{array}
\right.
\label{anom131}
\end{equation}
up to logarithmic prefactors depending on $\gamma_\rho$ and
$\gamma_t$.  Substituting
eqs.(\ref{anom129}), (\ref{anom130}), and (\ref{anom131}) in
eq.(\ref{anom113}), we get $\rho_{ALS}\sim\rho$. Thus, we have checked
that the ALS determined by the saddle-point solution (\ref{anom127})
indeed provides the value of LDOS which is equal (within the accuracy
of our consideration) to $\rho$. This confirms an assumption that an
anomalously large value of $\rho$ is typically governed by a single
ALS with the spatial structure described by the corresponding saddle
point configuration. Like in the quasi-1D case, the enhancement of
$\rho$ is determined by the product of three factors:
$|\psi^2(r)|_{\mbox{smooth}}$, $A(\rho)$, and $t_\phi/t_D$,
represented by eqs. (\ref{anom129}), (\ref{anom130}), and
(\ref{anom131}), respectively.

\subsubsection{Distribution of global density of states.}
\label{5.2.3}
Now, we consider the contribution of  ALS to the asymptotics of the
distribution of global DOS, which can be estimated according to
eq.(\ref{anom123}) as follows:
\begin{equation}
{\cal P}({\nu})\sim {\cal P}\left(t_\phi={\pi{\nu}\over
2\Delta\nu_{0}}\right)\sim
\left\{
\begin{array}{ll}
(g{\nu}/\nu_{0})^{-4\pi g}\ ,&\qquad {{\nu}\over\nu_{0}}\ll{1\over
g}\left({L\over l}\right)^2 \\
\displaystyle{ \exp\left\{-{\pi g\over
2}{\ln^2({\nu}/\nu_{0}\Delta\tau)\over\ln(L/l)}\right\}\ , }
&\qquad {{\nu}\over\nu_{0}}\gg{1\over
g}\left({L\over l}\right)^2
\end{array}
\right.
\label{anom132}
\end{equation}
The far LN asymptotic tail  in eq.(\ref{anom132}) is in full agreement
with the RG calculation by Altshuler, Kravtsov, and Lerner
\cite{akl}. We find also an intermediate power-law behavior, which
could not be obtained from the study of cumulants  in Ref.\cite{akl}. We
note, however, that this power-law form is fully consistent with the
change of the behavior of cumulants
$\langle\!\langle{\nu}^n\rangle\!\rangle$ at $n\sim\pi g$
discovered in \cite{akl}.

Finally, taking into account the close
similarity between the cumulants of DOS and of conductance \cite{akl},
it is natural to suppose that the distribution function of conductance
in a 2D metallic system has essentially the same behavior
(\ref{anom132}), with an intermediate power-law regime. This
hypothesis, which would be fully consistent with a power-law
asymptotic behavior of conductance distribution function on the
mobility edge suggested by Shapiro \cite{shapiro},
 needs however further verification.

\subsection{3D geometry.}
\label{s5.3}
As has been already mentioned in Sec.\ref{s4}, in three dimensions
the ALS are in
fact not localized. Their intensity just shows a relatively ``narrow''
bump on top of the usual average value $|\psi^2|=1/V$. Comparing the
results of Sec.\ref{s4} and of Refs.\cite{relax,ldos-as},
we find that the spatial
shape of these ``bumps'' is the same for the states, which are optimal
for all the distributions ${\cal P}(u)$, ${\cal P}(\rho)$, and $G(t)$.
Namely, it has the form
\begin{equation}
|\psi^2(r)|_{\mbox{smooth}}\sim{1\over V}\exp\left\{C_i{l\over r}\ln^2
Z_i\right\} \ ,\qquad i=u,\rho,t\ ,
\label{anom133}
\end{equation}
where
\begin{eqnarray}
Z_u&=&{uV\over (k_F l)^2}\ ,  \nonumber \\
Z_\rho&=&\rho/\nu_{0}\ ,\nonumber\\
Z_t& =&{t\over \tau (k_F l)^2}\ ,\nonumber
\end{eqnarray}
and $C_i$ are numerical constants of order of unity.
The corresponding distributions have one and the same,
log-cube-exponential, form:
\begin{equation}
{\cal P}(i)\sim \exp\{-\mbox{const}(k_F l)^2\ln^3 Z_i\}
\label{anom134}
\end{equation}
Note that formation of a large value of LDOS $\rho(E,r)$ cannot be
explained in 3D as a contribution of a single ALS. Indeed, in a
metallic sample LDOS is typically provided by a number of levels of
order of $g\sim k_F^2 lL$. In order that a single level might give
such (or even larger) a value of LDOS, it should has a local amplitude
(or, alternatively, an inverse width in energy space) enhanced by a
factor of $g$. However, in contrast to the quasi-1D and 2D
situations, $g$ does not enter the asymptotics (\ref{anom134}), which
do not depend on the system size $L$. Therefore, a high value of LDOS
is in 3D typically due to contribution of a large number ($\propto
L/l$) of adjacent levels.

\section{Additional comments.}
\label{s6}
\setcounter{equation}{0}

\subsection{States localized near the boundary.}
\label{s6.1}
We  assumed throughout the paper  that the center of an ALS is located
far enough from the sample edge. For a quasi-1D sample, this means
that $\xi_{ef}\ll L_+,L_-$. In the 2D case this implies that the
distance from the observation point to the boundary is of the same
order of magnitude in all directions, so that $\ln(L/l)$ is defined
without ambiguity. Here, we will consider briefly the role of ALS
situated close to the boundary, when these conditions are violated.

We start from the quasi-1D geometry. Let us calculate the distribution
function ${\cal P}(u)$ in a point located very close to one of the
sample edges. Formally, this means that $L_-\ll\xi_{ef}$. Then the
function $W^{(1)}(uS\xi,\tau_-)$ in eq.(\ref{anom11}) can be
approximated by unity, and we get
\begin{equation}
{\cal P}(u)={2\over \pi}\xi^{3/4} S^{1/4}L^{-1/2} u^{-3/4}
 \exp \left\{-2\sqrt{u\xi S}
+{\pi^2\xi\over 4L_+}\left(1-{\sqrt{\xi/ uS}\over L_+}+\ldots\right)
\right\}
\label{anom135}
\end{equation}
We see therefore that close to the boundary the distribution $\cal
P(u)$ has the asymptotic decay ${\cal P}(u)\sim\exp\{-2\sqrt{uS\xi}\}$,
which is slower than in the bulk of the sample,
${\cal P}(u)\sim\exp\{-4\sqrt{uS\xi}\}$. This means that if we consider
the distribution ${\cal P}(u)$ averaged over the position of the
observation point, its asymptotic tail will be always dominated by
contribution of the points located close to the boundary,
${\cal P}(u)\sim\exp\{-2\sqrt{uS\xi}\}$. This could be already anticipated
from eq.(\ref{anom50}), where the factor $\exp\left\{{\pi^2\over
4}\left({\xi\over L_+} + {\xi\over L_-}\right)\right\}$ strongly
increase with approaching one of the sample edges. The same tendency,
but in a weaker form, is observed in eqs.(\ref{adm15a}), (\ref{adm8a}),
(\ref{adm15b}). Calculating the average intensity
$\langle|\psi^2(r)|\rangle_u$ of the corresponding ALS, we find that
at $r>l$ eqs.(\ref{anom54}), (\ref{anom58}) retain their validity,
with an additional overall factor of 2. At small $r$,
eq.(\ref{anom48c}) is slightly modified:
\begin{equation}
 \langle|\psi^2(r)|\rangle_u=
\left({u\over\xi
S}\right)^{1/2}\left[1+\sqrt{uS\xi}k_d(r)\right]
\label{anom136}
\end{equation}

In 2D, we can consider a sample of the semicircular shape, with the
observation point located in the center of the diameter serving as a
boundary. The saddle-point solution then has exactly the same form
(\ref{anom87}), and the ALS intensity is still given by
eq.(\ref{anom89}), with an additional factor 2. The asymptotic form
of the distribution function ${\cal P}(u)$ gets an extra factor $1/2$ in
the exponent:
\begin{equation}
{\cal P}(u)\sim\exp\left\{-{\pi^2\nu_{0} D\over 2}
{\ln^2\left({V u\over 2\pi^2\nu_{0} D}\ln{L\over
l_*}\right)\over\ln(L/ l_*) } \right\}
\label{anom137}
\end{equation}
We expect this result to be applicable to any 2D sample of a
characteristic size $L$, with a smooth boundary and the observation
point taken in the vicinity of the boundary.

We see therefore, that, very generally, the probability of formation
of an ALS with the center in a given point is strongly enhanced (via
an extra factor $1/2$ in the exponent), if this point lies close to
the sample edge. This leads to the additional factor $1/2$ in the
exponent in the asymptotical form of the distributions ${\cal P}(u)$ and
${\cal P}(\rho)$ near the boundary.

\subsection{Orthogonal symmetry class.}
\label{s6.2}
All the considerations in this paper can be straightforwardly
generalized to the systems with unbroken time reversal invariance
(orthogonal symmetry class). The main results are as follows:
\begin{itemize}
\item[i)] all formulas for the average spatial density
$\langle|\psi^2(r)|\rangle$ in the metallic samples retain their
validity. In particular, in the quasi-1D case, eqs.(\ref{anom48b}),
(\ref{anom48c}),
(\ref{anom54}), (\ref{anom58}) hold with the same definition of
$\xi=2\pi\nu_{0} SD$. In the far localized tail, eq.(\ref{anom48a}), $\xi$
is replaced by $\xi/2$, which is just the conventional dependence of
the localization length on the symmetry of ensemble;
\item[ii)] in the expressions for the asymptotics of all distribution
functions, an extra factor $1/2$ appears in the exponent;
\item[iii)] GUE--type fluctuations are replaced by   the GOE-type
ones, where appropriate.
\end{itemize}

 \section{Summary.}
\label{s7}
\setcounter{equation}{0}

In this paper, we have studied the spatial structure of the
anomalously localized sates in weakly disordered samples. Such states
appear to govern the asymptotical behavior of the distribution function
${\cal P}(u)$ of local amplitudes of eigenfunctions. In the quasi-1D
geometry, an ALS has an effective localization length $\xi_{ef}$ much
shorter than the conventional one $\xi$. We were able to
calculate exactly the average intensity of such a state. The found
spatial distribution of the ALS intensity,
$\langle|\psi^2(r)|\rangle\propto 1/(r+\xi_{ef})^2$, turned out to be in
agreement with the form of the solution of the saddle-point equation of
Ref.\cite{falef}. Thus, the saddle-point configuration indeed
 describes the average intensity of an ALS, as was conjectured in
Refs.\cite{mk,falef}. This allowed us to describe the spatial
structure of ALS in 2D and 3D, as well. We have also studied the
fluctuations of the ALS intensity and found them to be essentially of
the same GUE type, as for a typical delocalized eigenstate.

Not only the asymptotic behavior of ${\cal P}(u)$, but also that of
distributions of other quantities can be governed by a kind of
ALS. Here are several important examples of such quantitites: inverse
participation ratio $I_2$, relaxation time $t_\phi$, local density of
states $\rho(E,r)$, global density of states ${\nu}(E)$. We have
found that the spatial structure of ALS relevant to the asymptotic
behavior of different distributions may be different. This is because
an ALS constitutes an optimal fluctuation for one of the above
quantities, and the form of this fluctuation depends on the specific
characteristic, which is to be optimized. Finally, we have discussed
interrelations between asymptotics of various distributions mentioned
above. In the quasi-1D and 2D cases, this allowed us to present a
comprehensive picture, which explains all the asymptotics as governed
by exponentially rare events of formation of ALS.

\section{Acknowledgments.}
The author is grateful to D.E.Khmelnitskii for discussions, and to
Y.V.Fyodorov for discussions  and critical reading of
the manuscript. This work was supported by SFB 195 der Deutschen
Forschungsgemeinschaft.

\section*{Appendix. Joint distribution function of eigenfunction
intensities in two different spatial points.}
\setcounter{equation}{0}
\renewcommand{\theequation}{A.\arabic{equation}}
In this Appendix, we write down the results for the joint distribution
function of intensities of an eigenfunction in two spatial points,
\begin{equation}
\cal
P(u,v;r)=\langle\delta(|\psi(0)^2|-u)\delta(|\psi(r)^2|-v)\rangle\ ,
\label{anoma1}
\end{equation}
for a quasi-1D system. For $l<r\ll\xi$, the function ${\cal P}(u,v)$ can
be restored from its moments, eq.(\ref{anom65}). The result is
obtained in the following form:
\begin{eqnarray}
{\cal P}(u,v;r)&=&{1\over \pi^2 V}{\partial^2\over\partial u\partial
v}{1\over\sqrt{uv}} \int_0^\infty d\nu\, \nu\sinh(\pi\nu)
e^{-\tau_1(1+\nu^2)/4}\int_0^1 dz
{1\over\sqrt{1-z}}W^{(1)}\left({v\xi\over
1-z},\tau_2\right)
\nonumber\\
&\times &K_{i\nu}\left(2\sqrt{{v\xi\over 1-z}}\right)
 {\partial\over\partial z}\left\{
{1\over\sqrt{z}}W^{(1)}\left({u\xi\over
z},\tau_-\right)K_{i\nu}\left(2\sqrt{{u\xi\over z}}\right)\right\}
\label{anoma2}
\end{eqnarray}
In the opposite case, $r<l$, the expression (\ref{anom79}) for the
moments has to be used, yielding
\begin{eqnarray}
{\cal P}(u,v;r)&=&{\xi\over L}{\partial \over \partial u}{\partial \over
\partial v }\int_0^{2\pi}{d\phi\over 2\pi}{\partial\over \partial
z}\{W^{(1)}(z,\tau_+) W^{(1)}(z,\tau_-) \}|_{z=z(u,v,r,\phi)}
\label{anoma3}\\
z(u,v,r,\phi)&=&\xi S\left({u\over 1+\sqrt{k_d(r)}e^{i\phi}} +
{v\over 1+\sqrt{k_d(r)}e^{-i\phi}}\right)
\nonumber
\end{eqnarray}
Unfortunately, even in the simplest case of an infinitely long sample,
when the function $W^{(1)}$ is given by eq.(\ref{anom38}),
formulas (\ref{anoma2}) and (\ref{anoma3}) are too involved. This is
why we have chosen in Sec.\ref{s3} to analyze  the expressions for
the moments, rather than the distribution function itself.


\begin{references}
\bibitem{note1} A similar idea was used to study the statistics of LDOS
and tunneling conductances, see Refs.\cite{epr,epri,mf5,mf6}.
\bibitem{note2} We are considering a quasi-1D sample, which means that
its length $L$ is much larger than the transverse size $a$. However,
$a$ is still supposed to be much larger than the wave length: $ak_F\gg
1$. On a scale shorter than $a$, the sample geometry is then not
of 1D, but rather of 2D or 3D nature, which is reflected in the
Friedel function (\ref{anom27}).
\end{references}
\end{document}